\documentclass[12pt]{article}

\pdfoutput=1
\usepackage[latin9]{inputenc}
\usepackage[top=80pt,bottom=85pt,left=67pt,right=67pt]{geometry}
\usepackage{amssymb,amsmath}
\usepackage{xypic,subfigure}
\usepackage{graphicx,color}
\usepackage{bbm}
\usepackage{float}
\usepackage{tensor}
\usepackage{cancel}
\usepackage{slashed}
\usepackage[debug,pageanchor=false]{hyperref}
\definecolor{link}{rgb}{.8,.15,.1}
\hypersetup{colorlinks=true,linkcolor=link,citecolor=link,urlcolor=link,linktocpage}

\usepackage[
  backend=biber,
  style=numeric-comp,
  sorting=none,
  sortcites=true,
  giveninits=true,
  maxbibnames=99,   
  maxcitenames=2,    
  doi=false
]{biblatex}
\hypersetup{
    colorlinks=true,
    urlcolor=blue,
}

\makeatletter
\@addtoreset{equation}{section}
\makeatother

\allowdisplaybreaks

\newcommand{\beq}{\begin{equation}}
\newcommand{\eeq}{\end{equation}}
\newcommand{\bea}{\begin{eqnarray}}
\newcommand{\eea}{\end{eqnarray}}
\newcommand{\nn}{\nonumber}

\newcommand{\eq}{\begin{equation}}
\newcommand{\feq}{\end{equation}}
\newcommand{\eqn}{\begin{eqnarray}}
\newcommand{\feqn}{\end{eqnarray}}

\begin{document}

\begin{titlepage}

\begin{center}

\vskip .5in 
\noindent

{\Large \bf{M5 branes wrapping $\mathbb{WCP}^2$ and spindles fibred over constant curvature Riemann surfaces}}

\bigskip\medskip

Andrea Conti\footnote{contiandrea@uniovi.es}, Niall T. Macpherson\footnote{macphersonniall@uniovi.es}, Diego de Maria Almazan\footnote{mariadiego@uniovi.es}\\

\bigskip\medskip
{\small 

Department of Physics, University of Oviedo,
C/ Leopoldo Calvo Sotelo, 18, 33007 Oviedo, Asturias}

\medskip
{\small and}

\medskip
{\small 

Instituto Universitario de Ciencias y Tecnolog\'ias Espaciales de Asturias (ICTEA),\\
Calle de la Independencia 13, 33004 Oviedo, Spain}\\

\medskip

\vskip 2cm 

     	{\bf Abstract }\\[2mm]
			     	\end{center}
We classify AdS$_3$ solutions of the U(1) invariant sector of minimal $d=7$ supergravity. We find two classes of solutions preserving ${\cal N}=(2,0)$ supersymmetry for which the internal space M$_4$ is either a negative curvature Kahler-Einstein manifold or a circle fibration over $\Sigma\times \mathbb{R}$. For the later, in the case that $\Sigma$ has constant curvature, we reduce finding a solution to solving a single ODE that admits polynomial solutions. Among these are interesting solutions whose uplifts to $d=11$ describe M5 branes wrapping various $d=4$ orbifolds. These include a topological $\mathbb{CP}^2$ with 2 orbifold fixed points that we identify as the weighted projective space $\mathbb{WCP}^2_{[k,k,\ell]}$. We are also able to construct solutions with M5 branes that wrap a spindle fibred over constant curvature Riemann surfaces of arbitrary genus. Such solutions should provide holographic duals to the ${\cal N}=(2,0)$ SCFT associated to the M5 brane compactified to $d=2$ on these orbifolds. We match the holographic central charges of these solutions to a field theory computation in terms of anomaly polynomials and c-extremisation. 

     	\noindent

\noindent

\vfill
\eject

\end{titlepage}

\tableofcontents

\section{Introduction}

Starting with \cite{Maldacena:2000mw}, an area of the AdS/CFT correspondence that has seen much interplay between the geometry and CFT sides has been backgrounds describing D3, M2 and M5 branes wrapped on various compact $k$-cycles, and their dual descriptions in terms of CFTs compactified on these cycles. On the CFT side this allows one to start with a maximally supersymmetric theory in $d$ dimensions, such as  ${\cal N}=4$ super Yang-Mills in $d=4$ or the ${\cal N}=(2,0)$ SCFT associated to the M5 brane in $d=6$, and systematically construct wide classes of theories in $d-k$ dimensions that preserve some portion of the original supersymmetry. In its original formulation this process necessitates that the CFT$_d$ is topologically twisted \cite{Witten:1988ze,Bershadsky:1995qy}, which involves a coupling to both metric and gauge fields on the $k$-cycle such that supersymmetry is preserved on the, now curved, $d$ dimensional space\footnote{This involves identifying the R-symmetry gauge field with the spin connection on the $k$-cycle such that the supported spinors are still constant, as they would be in flat space, and that the Chern class of the gauged field is equal to the Euler characteristic of the $k$-cycle.}. In many cases a similar prescription can be followed on the geometry side to construct the AdS$_{d+1-k}$ solutions that are dual to these SCFTs. This utilises the $d+1$ dimensional gauged supergravities that arise upon consistent truncation about the near horizon geometry of the brane in question. The gauge fields in the CFT and supergravity are essentially identified.

The focus of \cite{Maldacena:2000mw} was branes wrapping $\Sigma_{\mathfrak{g}}$, a compact constant curvature Riemann surface of genus $\mathfrak{g}$, and SCFTs compatified on this surface. Since this work, there have been many others that construct AdS$_{p}\times \Sigma_{\mathfrak{g}}$ solutions such as \cite{Brinne:2000fh,Gauntlett:2001qs,Bah:2011vv,Anderson:2011cz,Bah:2012dg,Bah:2013qya,Benini:2013cda,Apruzzi:2015zna,Couzens:2021tnv,Benini:2015bwz}, as well as solutions describing branes wrapping higher dimensional constant curvature surfaces \cite{Gauntlett:2000ng,Nunez:2001pt,Donos:2010ax,Benini:2013cda,Rota:2015aoa}, and non-constant Riemann surfaces with punctures \cite{Gaiotto:2009gz,Bah:2017wxp,Bobev:2019ore,Bah:2019jts,Apruzzi:2025znw}. A commonality among these works is that supersymmetry is preserved in terms of a topological twist.

Beginning with \cite{Ferrero:2020laf}, it was realised that one could also wrap branes on spindles, which are Riemann surfaces of non-constant curvature that also preserve supersymmetry. Specifically the spindle is the weighted projective space $\mathbb{WCP}^{1}_{[k_1,k_2]}$ which has the topology of a 2-sphere with $\mathbb{R}^2/\mathbb{Z}_{k_{1,2}}$ orbifold fixed points at the respective poles. A novel aspect of such constructions is that supersymmetry is not preserved in terms of a topological twist, yet precise matching of geometric and CFT quantities can still be made. Following \cite{Ferrero:2020laf}, there have appeared diverse works containing spindles such as \cite{Ferrero:2020twa,Ferrero:2021etw,Ferrero:2021ovq,Ferrero:2021wvk,Hosseini:2021fge,Boido:2021szx,Couzens:2021rlk,Couzens:2021cpk,Arav:2022lzo,Boido:2022mbe,Couzens:2022yiv,Suh:2023xse,Inglese:2023wky,Inglese:2023tyc,Hristov:2023rel,Amariti:2023mpg,Pittelli:2024ugf,Amariti:2023gcx,Hosseini:2023ewi,Bomans:2024mrf,Colombo:2024mts,Hristov:2024qiy,Crisafio:2024fyc,Ferrero:2024vmz,Boisvert:2024jrl,Arav:2025jee, Jeon:2025rfc,Suh:2025cii,Conti:2025rfd,Jeon:2026kxn,Arav:2026unc} which are not limited to wrapped brane constructions but include scenarios with spindles appearing in black hole near horizon geometries and within a field theory localisation context.

It is natural to wonder whether supersymmetry is also preserved when branes wrap higher dimensional orbifolds of non-constant curvature, indeed there have been several works that address this. In \cite{Cheung:2022ilc} solutions with M5 branes that wrap a spindle fibred over either $\Sigma_{\mathfrak{g}}$ for $\mathfrak{g}>1$ or a second spindle were constructed, and their SCFT$_2$ duals considered. The method employed to achieve this was to construct a consistent truncation about an AdS$_5\times \mathbb{WCP}^{1}_{[k_1,k_2]}$ solution in maximal $d=7$ gauged supergravity \cite{Ferrero:2021wvk}, which itself admits an uplift to $d=11$. The truncation was to minimal $d=5$ gauged supergravity which contains supersymmetric AdS$_3$ solutions whose internal space is either a spindle or $\Sigma_{\mathfrak{g}}$. See also \cite{Couzens:2022lvg,Suh:2022olh,Suh:2024fru} which follow a similar path to \cite{Cheung:2022ilc}. Another notable work is \cite{Faedo:2024upq}, which considered D4 and M5 branes wrapping ``quadrilaterals'', 4d toric orbifolds with 4 fixed points. These solutions where constructed via double analytic continuation of certain black hole/string near horizon geometries \cite{Chow:2011fh},\footnote{More properly, a Nutt charge is added to these black holes first.} which was also the method employed to construct many of the spindle solutions. 

The most natural higher dimensional generalisation of the spindle is $\mathbb{WCP}^{2}_{[k_1,k_2,k_3]}$, which is to the spindle as $\mathbb{CP}^2$ is to the 2-sphere. As the notation suggests, this space is topologically $\mathbb{CP}^{2}$ with 3 orbifold singularities associated to the integer weights $k_i$. Within a supergravity context $\mathbb{WCP}^{2}_{[1,1,2]}$ has appeared in \cite{Gauntlett:2004zh,Bianchi:2021uhn},  $\mathbb{WCP}^{2}_{[k_1,k_1-k_2,k_2]}$ in \cite{Macpherson:2024frt}, and generic $\mathbb{WCP}^{2}$ and $\mathbb{WCP}^{3}$ in \cite{Conti:2025rfd} - see also \cite{Mauch:2024uyt} for a localisation perspective. An area where $\mathbb{WCP}^{2}$ is notably absent from the literature is as a manifold that branes are wrapped on. This is a shame because it is a promising non-constant curvature manifold on which to compactify the world volume of M5 branes, and the (2,0) SCFT associated to them, to 2 dimensions. Constructing such a supersymmetry preserving AdS$_3$ solution is the main motivation of this work.

In all known examples, one of the orbifold singularities in $\mathbb{WCP}^{2}$ is $\mathbb{R}^4/\mathbb{Z}_k$, which involves the entire manifold shrinking to zero. As such it is hard to see how one could construct this orbifold from the sort of spindle compactification of \cite{Cheung:2022ilc}, or some generalisation there of. Likewise, while many spindle solutions were constructed from double analytic continuations of black holes, there is no particularly good reason to expect $\mathbb{WCP}^{2}$ to arise easily in the same fashion. So how to construct M5 branes wrapping $\mathbb{WCP}^{2}$? As the generic example in \cite{Conti:2025rfd} is a U(1) orbifold fibration over $\mathbb{WCP}^{1}$, one approach might be to invert the logic of \cite{Cheung:2022ilc} and attempt to lift the AdS$_3\times\mathbb{WCP}^{1}$ solution of  \cite{Ferrero:2020laf} to $d=7$ maximal gauged supergravity using G-structures. Likely this would require one to classify the possible 2d embedding manifolds which permit an uplift in a similar fashion to \cite{Macpherson:2025sff}, then attempt to find one that lifts $\mathbb{WCP}^{1}$ to $\mathbb{WCP}^{2}$. Here we will follow a different path: Instead we will classify the AdS$_3$ solutions of the U(1) invariant sector of minimal $d=7$ gauged supergravity, which admits uplifts to both $d=11$ and massive IIA, and leverage this result to construct solutions with M5 branes wrapping a particular version of $\mathbb{WCP}^{2}$. As a bi-product we will also be able to construct solutions that describe M5 branes wrapping a spindle fibred over totally general compact Riemann surfaces\footnote{Which is to say not only are we compatible with $\mathfrak{g}=0,1$ and $\mathfrak{g}>1$, but in addition for us $\mathfrak{g}>1$ is otherwise unconstrained while in  \cite{Cheung:2022ilc} it must be tuned to be a combination of the weights on the spindle.}.\\
~\\

The lay out of the paper is as follows:\\

In section \ref{sec:minimalgaguedsupergravity} we review minimal gauged supergravity in 7 dimensions, its U(1) invariant sub-sector and its embedding into $d=11$ supergravity, in terms of S$^4/\mathbb{Z}_{\tilde{k}}$, and massive IIA.

In section \ref{eq:AdS3classificaion} we classify the supersymmetric AdS$_3$ solutions of the U(1) sector of $d=7$ minimal gauged supergravity in terms of bi-spinors and the G-structure forms that span them\footnote{A partial classification of more generic solutions of this theory was performed in \cite{Cariglia:2004qi}.}. We find two classes of solutions for which the internal manifold M$_4$ takes a distinct form: First in section \ref{eq:sec:KEclass} we derive a class where M$_4$ is a negative curvature Kahler-Einstein manifold that generalises a solution from \cite{Gauntlett:2000ng} to also include a non-trivial point dependent scalar. The second class of section \ref{sec:class2} is more interesting for our purposes, in it M$_4$ supports an identity-structure. We show that M$_4$ decomposes as a U(1) fibration over a Riemann surface that is foliated over an interval. We reduce finding a solution in the class to solving 3 PDEs on M$_4$. 

Section \ref{sec:CC} takes the class of \ref{sec:class2} and imposes that the Riemann surface it contains has constant curvature. After this simplification, solutions are in one to one correspondence with a single second order ODE.  We are able to find 3 polynomial solutions to this ODE, 2 of which contain physical solutions. Among these are several known solutions from the literature we recover in appendix \ref{sec:knownsols}. One of the polynomial solutions admits a non-trivial gauge field and many distinct global completions - studying those for which M$_4$ is compact and the CFTs they are dual to constitutes the rest of the paper.

In section \ref{eq:M5braneswrappingWCP2} we construct an  AdS$_3\times \mathbb{WCP}^2$ solution whose uplift to $d=11$ provides the first example of M5 branes wrapping $\mathbb{WCP}^2$. The specific 4 dimensional orbifold we find is $\mathbb{WCP}^2_{[k,k,\ell]}$, where  $\ell$ needs to be tuned to realise a well defined fibration. A well defined solution in $d=11$ requires one to either further tune $\ell$ or $\tilde{k}$ in  S$^4/\mathbb{Z}_{\tilde{k}}$.

We present the remaining solutions with  compact M$_4$ in section \ref{sec:M5spindlesSigma}. For these the internal space takes the form of a $\mathbb{WCP}^1_{[k_1,k_2]}$ fibration over $\Sigma_{\mathfrak{g}}$, a constant curvature Riemann surface of arbitrary genus. Their embeddings into $d=11$ require a tuning of either $(k_1,k_2)$ or $\tilde{k}$ to have a well defined solution. 

In section \ref{eq:fieldtheory} we consider the 2 dimensional CFTs dual to our AdS$_3\times\text{M}_4$ solutions. Via a field theory computation utilising the M5 brane anomaly polynomial (see for instance \cite{Witten:1996hc,Alday:2009qq,Hosseini:2020vgl}) and c-extremisation \cite{Benini:2012cz}, in the large $N$ limit, we are able to precisely match the holographic central charge and R-symmetry of our geometries. 

Finally we make some concluding remarks in section \ref{sec:conclusions}. The work is supplemented by several appendices referred to in the main text.

\section{Minimal $d=7$ gauged supergravity and its U(1) invariant sector}\label{sec:minimalgaguedsupergravity} 
In this section we review minimal $d=7$ gauged supergravity, its U(1) invariant sub-sector and its embeddings into $d=11$ and massive type IIA supergravities.\\ 
~\\
Minimal $d=7$ gauged supergravity \cite{Townsend:1983kk}\footnote{We follow the conventions of \cite{Passias:2015gya}.} contains only a gravity multiplet whose bosonic constituents are a metric $g_{\mu\nu}$, a triplet of SU(2) gauge fields $A_{\mu}^i$, a scalar $\varphi$ and a 4-form ${\cal F}_4$. Upon introducing a potential ${\cal A}_3$, we can define field strengths as
\beq
{\cal{F}}^i = d {\cal{A}}^i - \frac{1}{2} g \epsilon^{ijk} {\cal{A}}^j \wedge {\cal{A}}^k,~~~~ {\cal{F}}_4 = d {\cal{A}}_3,
\eeq
which are such that
\beq
D{\cal F}^i=0,~~~~d{\cal F}_4=0,
\eeq
where $D$ denotes the gauge-covariant exterior derivative $ D \omega^i = d \omega^i - g \, \epsilon_{ijk} A^{j} \wedge \omega^k $. The pseudo-action of the bosonic theory  is then given in terms of these by
\begin{align}
S=& \int\bigg[\star_7 (R-V)-\frac{1}{2}\star d\varphi\wedge d\varphi-\frac{1}{2}X^4\star_7{\cal F}_4\wedge {\cal F}_4-\frac{1}{2}X^{-2}\star_7 {\cal F}^i\wedge {\cal F}^i\nn\\[2mm]&+\frac{1}{2}{\cal F}^i\wedge {\cal F}^i\wedge {\cal A}_3-h {\cal F}_4\wedge {\cal A}_3\bigg],\label{eq:action}
\end{align}
where we have defined 
\beq
X=e^{\frac{1}{\sqrt{10}}\varphi},~~~~V=2h^2 X^{-8}-4\sqrt{h}g X^{-3}-2g^2 X^2,
\eeq
where $g$ is the gauge coupling and $h$ is a ``topological mass''.  
The above needs to be supplemented by the ``odd-dimensional self-duality" equation
\begin{equation}\label{oddselfduality}
X^4 * \mathcal{F}_4 = - \frac{1}{\sqrt{2}} g \mathcal{A}_3 + \frac{1}{2} \omega_3,
\end{equation} 
where $\omega_3 \equiv {\cal{A}}^i \wedge {\cal{F}}^i + \frac{1}{6} g \epsilon^{ijk} {\cal{A}}^i \wedge {\cal{A}}^j \wedge {\cal{A}}^k$, satisfying $d\omega_3 = {\cal{F}}^i \wedge {\cal{F}}^i$ - Note that on shell this essentially has no consequence as it just serves as a definition for ${\cal A}_3$.
  
When $\frac{h}{g}>0$ the theory admits a supersymmetric AdS$_7$ vacuum and we can elect to fix
\beq
h=\frac{1}{2\sqrt{2}}g,
\eeq 
without loss of generality. This assumption allows us to make contact with supersymmetry preserving consistent truncations about massive IIA and $d=11$ supergravity. Given this  and \eqref{oddselfduality}, the action \eqref{eq:action} then  gives rise to the following equations of motion for the fluxes
\begin{subequations}
\begin{align}
d(X^4* {\cal{F}}_4) + \frac{1}{\sqrt{2}} g \mathcal{F}_4 - \frac{1}{2}  {\cal F}^i \wedge {\cal F}^i=0, \label{7dF2}\\[2mm]
D(X^{-2}*   {\cal F}^i) -  {\cal F}^i \wedge  {\cal F}_4=0 \label{7dF4},
\end{align}
\end{subequations}
while Einstein's equations and the scalar equation of motion take the form
\begin{align}
&R_{\mu\nu} - 5 X^{-2} \partial_\mu X \partial_\nu X - \frac{1}{20} g^2 \left(X^{-8} - 8 X^{-3} - 8 X^2\right) g_{\mu\nu} - \frac{1}{2} X^{-2} \left( ({{\cal{F}}^{i2}})_{\mu\nu} - \frac{1}{5} {{\cal{F}}^i}^2 g_{\mu\nu}\right)\nn \\[2mm]
& - \frac{1}{2}X^4\left( ({{\cal{F}}_4}^2)_{\mu \nu} - \frac{3}{5}  {\cal F}_4^2 g_{\mu\nu}\right) = 0,\label{7dEinsteinEquations}\\[2mm]
& d(X^{-1}*_7dX) + \frac{1}{5} g^2(X^{-8}-3X^{-3}+2 X^2)\text{vol}_7 - \frac{1}{5}X^{4} *_7  {\cal{F}}_4 \wedge {\cal{F}}_4 + \frac{1}{10} X^{-2} *_7 {\cal{F}}^i \wedge {\cal{F}}^i=0,\nn
\end{align} 
where in the above we define
\begin{align}
({\cal{F}}^2)_{MN} & \equiv \frac{1}{(p-1)!}\sum_p {{\cal{F}}_p}_M{}^{M_1 \dots M_{(p-1)}} {{\cal{F}}_p}_{N M_1 \dots M_{(p-1)}}, \\[2mm]
{\cal F}^2_p & \equiv \frac{1}{p!}\sum_p {{\cal{F}}_p}^{M_1 \dots M_{p}} {{\cal{F}}_p}_{M_1 \dots M_{p}}.
\end{align}
Together these conditions define solutions to the theory.

A solution is supersymmetric if there exists a pair of non-trivial symplectic Majorana spinors $\zeta_a$ for $a=1,2$ obeying the conditions
\begin{align}
&(\nabla_{\mu}+i g({\cal A})_a^{~b})\zeta_b+\frac{i}{10\sqrt{2}}X^{-1}(\gamma_{\mu}^{~\alpha\beta}-8 \delta_{\mu}^{\alpha}\gamma^{\beta})({\cal F}_{\alpha\beta})_{a}^{~b}\zeta_b\nn\\[2mm]
&+ \frac{1}{160}X^2(\gamma_{\mu}^{~\alpha_1\alpha_2\alpha_3\alpha_4}-\frac{8}{3}\delta_{\mu}^{~\alpha_1}\gamma^{\alpha_2\alpha_3\alpha_4}{\cal F}_{4\alpha_1\alpha_2\alpha_3\alpha_4})\zeta_a
-\frac{g}{20\sqrt{2}}(4X+X^{-4}) \gamma_{\mu}\zeta_a=0,\nn\\[2mm]
&\frac{5}{2}X^{-1}dX\zeta_a-\frac{i}{\sqrt{2}}X^{-1}({\cal F})_{a}^{~b}\zeta_b+\frac{1}{2}X^2{\cal F}_4\zeta_a-\frac{g}{\sqrt{2}}(X^{-4}-X)\zeta_a=0,
\end{align}
where we define
\beq
{\cal A}_{a}^{~b}= \frac{1}{2}{\cal A}^{i}(\sigma^i)_a^{~b},~~~~{\cal F}_{a}^{~b}= \frac{1}{2}{\cal F}^{i}(\sigma^i)_a^{~b},
\eeq
and a $k$-form $C_k$ acts on a spinor as
\beq
C_k\zeta=  \frac{1}{k!}(C_k)_{\alpha_1...\alpha_k}\gamma^{\alpha_1...\alpha_k}\zeta.
\eeq

\subsection{The U(1) invariant sector}
In this work we will be interested in the U(1) invariant sector of minimal $d=7$ gauged supergravity. This can be reached by fixing
\beq
{\cal A}^1={\cal A}^2=0,~~~~~{\cal A}^3={\cal A},~~~~~{\cal F}=d{\cal A},
\eeq
such that ${\cal F}$ is closed. This has the effect of sending $D\to d$ in the equations of motion. In this case, the conditions for supersymmetry can also more usefully, at least for our proposes, be expressed in terms of a single Dirac spinor $\zeta$ as
\begin{align} 
& (\nabla_\mu + \frac{i}{2} g {\cal{A}}_{\mu})\zeta+ \frac{i}{20 \sqrt{2}} X^{-1}\left(  5{\cal{F}} \gamma_{\mu}  - 3 \gamma_{\mu} {\cal{F}} \right)\zeta + \frac{1}{40} X^2 \left( \gamma_{\mu} {\cal{F}}_4  + 5{\cal{F}}_4 \gamma_{\mu} \right) \zeta - \frac{g}{20\sqrt{2}}\left( X^{-4} + 4 X\right)\gamma_\mu \zeta=0,\nn\\[2mm]
&\bigg[\frac{5}{2}X^{-1}dX-\frac{i}{2\sqrt{2}}X^{-1}{\cal F}+\frac{1}{2}X^2{\cal F}_4-\frac{g}{\sqrt{2}}(X^{-4}-X)\bigg]\zeta=0,\label{eq:susy7dU1}
\end{align}
where the symplectic Majorana spinors $\zeta_a$ is defined in terms of this as $\zeta^a=(\zeta,\zeta^c)^T$.

\subsection{Embeddings into $d=10$ and $d=11$}\label{eq:1011duplifts}
In this work, for simplicities sake and to make contact with wrapped M5 branes, we will be mainly concerned with embeddings of $d=7$ gauged supergravity into $d=11$ supergravity: As we derive in appendix \ref{sec:uplift11d}, for the U(1) invariant sector the $d=7$ fields are embedded in the $d=11$  metric as
\begin{align}
\frac{ds^2}{L^2}& =  g^2\Delta^{1/3}g_{\mu\nu}dx^{\mu}dx^{\nu}+2\Delta^{-2/3}\bigg[ X^3\Delta d\beta^2+\frac{\cos^2{\beta}}{X}\left(\frac{1}{4}Ds^2(\text{S}^2)+\frac{1}{\tilde{k}^2}D\hat{\psi}^2\right)\bigg],\nn\\[2mm]
\Delta&= X^{-4}\sin^2\beta+X\cos^2\beta,~~~~Ds^2(\text{S}^2)=d\theta^2 + \sin^2\theta D\phi^2, \label{eq:11dmet}
\end{align}
where the covariant exterior derivatives take the form
\begin{equation}
D\phi=d\phi - g {\cal{A}},~~~D\hat{\psi}=d \hat{\psi}- \frac{\tilde{k}}{2}\cos\theta D\phi.
\end{equation}
The $d=7$ fields also appear in the 4-form flux as
\begin{align}
\frac{G}{L^3} & = \frac{g}{\sqrt{2}} \bigg( \sin^2\theta\cos\beta d\beta\wedge\mathcal{F}\wedge D\phi -\frac{\cos^2\beta\sin\beta}{2 X^4 \Delta}\cos\theta\mathcal{F}\wedge\text{vol}_2 -\sqrt{2} g^2 \sin\beta\mathcal{F}_4 \nn\\[2mm]
& + 2 g X^4 \cos\beta d\beta\wedge\star_7\mathcal{F}_4 \bigg) + \frac{\sqrt{2}}{\tilde{k}} D\hat{\psi}\wedge\bigg(g\cos\beta\cos\theta d\beta\wedge\mathcal{F} \\[2mm]
& + \frac{g\cos^2\beta\sin\beta}{2 X^4 \Delta} \sin\theta d\theta \wedge \mathcal{F} - \frac{5\cos^4\beta\sin\beta }{2 X^4 \Delta^2} dX \wedge \text{vol}_2 + \frac{(2X^5-1) \Delta + 2 X }{2 X^4 \Delta^2} \cos^3\beta d\beta\wedge\text{vol}_{2} \bigg).\nn
\end{align}
where $\text{vol}_{2}= \sin\theta d\theta\wedge D\phi$. Although it is not totally clear from the above expressions, the fibred topological 4-sphere can actually be decomposed in terms of a round 3-sphere with ${\cal A}$ only appearing in the U(1) of the Hopf fibre direction, parametrised by $\partial_{\phi}$ \cite{Lu:1999bc}. While such a parametrisation might appear simpler it is important to appreciate that when a supersymmetric $d=7$ solution is embedded into $d=11$, it is the $\hat{\psi}$ coordinate and not $\phi$ that we are free to reduce to IIA on without breaking supersymmetry entirely. For similar reasons we can take a supersymmetry preserving $\mathbb{Z}_{\tilde{k}}$ orbifold of the $\hat{\psi}$ direction. \\
~\\
The above of course intersects, an orbifold of, the embedding of maximal $d=7$ gauged supergravity into $d=11$ \cite{Pilch:1984xy}. Indeed, in the conventions of \cite{Cheung:2022ilc}, the fields of this U(1)$^2$ invariant sector, considered in \cite{Cvetic:1999xp}, are embedded into the $d=11$ metric as\footnote{We have exploited trombone symmetry to rescale the metric in appendix A.2 of \cite{Cheung:2022ilc} by $2L^2 g_c^2$ and have set $(A^{11}_{(1)}=A_1,~A^{22}_{(1)}=A_2)$.} 
\begin{align}
\frac{ds^2}{L^2}&= 2\Delta^{\frac{1}{3}} g_c^2ds^2_7+2 \Delta^{-\frac{2}{3}}\bigg[e^{4(\lambda_1+\lambda_2)}dw_0^2+\sum_{a=1}^2e^{-2\lambda_a}\left(dw_a^2+w_a^2(d\chi_a-g_c A_a)^2\right)\bigg],\nn\\[2mm]
\Delta&=  e^{-4(\lambda_1+\lambda_2)}w_0^2+ e^{2\lambda_1}w_1^2+ e^{2\lambda_2}w_2^2,~~~~w_0^2+w_1^2+w_2^2=1,\label{eq:maximalmet}
\end{align}
One can implement a supersymmetry preserving $\mathbb{Z}_{\tilde{k}}$ orbifold of the internal 4-sphere by identifying
\beq
\chi_1\sim \chi_1- \frac{2\pi}{\tilde{k}},~~~~\chi_2\sim \chi_2+ \frac{2\pi}{\tilde{k}}.
\eeq 
This action can be realised in terms of a pair of period $2\pi$ isometry directions $(\phi,\hat\psi)$ as 
\beq
\chi_1= \phi-\frac{1}{\tilde{k}}\hat\psi,~~~~\chi_2=\frac{1}{\tilde{k}}\hat\psi,
\eeq
which should of course be identified with the U(1)$^2$ isometry in \eqref{eq:maximalmet}. The limit in which the U(1)$^2$ sector truncates to minimal $d=7$ gauged supergravity is
\beq
A_1=A_2,~~~~ e^{\lambda_1}=e^{\lambda_2},
\eeq
and it is a simple matter to establish that one should identify
\beq
g_{c}= \frac{1}{\sqrt{2}}g,~~~~ A_1=A_2= \frac{1}{\sqrt{2}} {\cal A}, ~~~~e^{\lambda_1}=e^{\lambda_2}=X,~~~~ds^2_7=g_{\mu\nu}dx^{\mu}dx^{\nu} \label{eq:identifications}
\eeq
to match our conventions for the minimal theory. Indeed one can show that the above identifications precisely map \eqref{eq:maximalmet} to \eqref{eq:11dmet} upon taking
\beq
w_0=\sin\beta,~~~~w_1+i w_2= \cos\beta e^{i \frac{\theta}{2}}.
\eeq
This mapping allows us to extract some important information. First off, with respect to \eqref{eq:maximalmet}, when $A_a$ are non-trivial they define a fibration over the base $ds^2_7$. As such we should have that $g_c d{\cal A}_a$ are appropriately quantised over any  2-cycles in $ds^2_7$ they have support on such that this fibration is well defined. For the S$^4/\mathbb{Z}_k$ orbifold it then clearly becomes  $ g_c \tilde{k} d{\cal A}_a$ that must be quantised. The identifications in \eqref{eq:identifications} imply that $2 g_c A_1= 2 g_c A_2= g {\cal A}$, as such we should have that 
\beq
\frac{1}{2\pi}\int_{\Sigma_2}\frac{g \tilde{k}}{2} d{\cal A},\label{eq:quantisationcondtion}
\eeq
is appropriately quantised, for $\Sigma_2$ any 2-cycle in the $d=7$ geometry, \textit{i.e.} for $\Sigma_2=\text{S}^2$ the integral should be integer. Second, \cite{Cheung:2022ilc} identify the R-symmetry gauge field in \eqref{eq:maximalmet} as $A_R= g_c(A_1+A_2)=g {\cal A}$, which means that $g{\cal A}$ is the R-symmetry gauge field in \eqref{eq:11dmet}.\\
~\\
There is a more general embedding of the entire $d=7$ minimal theory into massive IIA which was derived in \cite{Passias:2015gya}. In terms of the parametrisation of AdS$_7$ solutions in \cite{Cremonesi:2015bld}, this intersects the $\partial_{\hat\psi}$ reduction of the $d=11$ embedding only when the Romans mass is zero. For the U(1) invariant sector, and in the presence of a non-trivial Romans mass, this takes the form
\begin{align}
\frac{ds^2}{\sqrt{2\pi}} & =g^2\sqrt{-\frac{\alpha}{\alpha''}}X^{-1/2}g_{\mu\nu}dx^{\mu}dx^{\mu} + X^{5/2}\sqrt{-\frac{\alpha''}{\alpha}}\left(dz^2 + \frac{\alpha^2}{\alpha'^2 - 2\alpha\alpha'' X^5} Ds^2(\text{S}^2)\right), \nn\\[2mm]
e^{\Phi} & = 3^4 2^{5/4} \pi^2 \sqrt{\pi} \frac{ X^{5/4}}{ \sqrt{\alpha'^2 - 2 X^5 \alpha  \alpha''}} \left(-\frac{\alpha}{\alpha''}\right)^{3/4},~~~
B_2  = -\pi \left(\frac{ \alpha \alpha' }{\alpha '^2 - 2 X^5 \alpha \alpha''} \text{vol}_2- \cos \theta dz \wedge D\phi\right),\nn\\[2mm]
F_0 & = -\frac{\alpha'''}{2\ 3^4 \pi ^3}, ~~~F_2  = \left(\frac{\alpha''}{3^4 2 \pi^2} + F_0 \frac{\pi \alpha \alpha' }{ \alpha'^2 - 2 X^5 \alpha \alpha''} \right) \text{vol}_2 + \frac{\alpha''}{ 3^4 2 \pi^2 } g \cos \theta {\cal{F}}, \\[2mm]
F_4 & = -\frac{g }{2 \pi  3^4} \left(\sin^2 \theta \alpha'' {\cal{F}} \wedge D\phi \wedge dz - \frac{\alpha \alpha' \alpha'' \cos \theta  }{\alpha'^2 - 2 X^5 \alpha \alpha''} {\cal{F}} \wedge \text{vol}_2 \right) - \frac{ g^3 \alpha'}{\sqrt{2} \pi 3^4} {\cal{F}}_4 + \frac{g^2 X^4 \alpha'' }{\pi 3^4} dz \wedge \star_7 {\cal{F}}_4\nn
\end{align} 
where the NS 3-form is $H=dB_2$. Solutions are governed by the, piece-wise, polynomial solutions to
\beq
\alpha(z)'''=-2\pi^3 3^4 F_0,
\eeq
where $F_0$ is the Romans mass. Although we will not discuss them explicitly, the above provides infinitely many embeddings of the $d=7$ solutions we later construct into massive IIA. Specifically, an embedding exists for each of the infinite family of globally distinct AdS$_7$ solutions in massive IIA.

\section{Classifying AdS$_3$ solutions}\label{eq:AdS3classificaion} 
In this section we classify supersymmetric AdS$_3$ solutions of the U(1) invariant sector of minimal $d=7$ supergravity in terms of G-structures.\\
~\\
We begin by making the following ansatz for the $d=7$ bosonic fields 
\begin{align}
ds^2 & = e^{2A} ds^2(\text{AdS}_3) + ds^2(\text{M}_4), \\[2mm]
{\cal{F}}_4 & = e^{3 A} \text{vol}(\text{AdS}_3) \wedge g_1 + g_4,
\end{align}
where the  AdS$_3$ warp factor $e^{2A}$, the scalar $X$ and the 1- and 4-forms $(g_1,g_4)$ have support on M$_4$ only, which is independentent of the AdS$_3$ directions. As such, the SO(2,2) symmetry of AdS$_3$ is preserved.
 
The $d=7$ Bianchi identities decompose into
\beq
d{\cal F}=0,~~~~d(e^{3A}g_1)=0,~~~~~dg_4=0,
\eeq
the equations of motion for the $d=7$ fluxes become
\begin{subequations}
\begin{align}
& d ( e^{3A} X^{-2} \star_4 {\cal{F}}) + e^{3A} g_1 \wedge {\cal{F}}= 0,\label{fluxeom1} \\[2mm]
& d (e^{3A} X^4 \star_4 g_4) - \frac{g}{\sqrt{2}} e^{3 A} g_1 = 0,\label{fluxeom2}\\[2mm]
& d \left( X^4 \star_4 g_1 \right) +  \frac{g}{\sqrt{2}} g_4 - \frac{1}{2} {\cal{F}} \wedge {\cal{F}} = 0, \label{fluxeom3}
\end{align}
\end{subequations}
and the equation of motion of the scalar reduces to
\begin{align}
&d \left( \frac{e^{3A}}{X} \star_4 dX \right) + \frac{e^{3A}}{5} X^4 g_1 \wedge \star_4 g_1 -\frac{e^{3A}}{10 X^2} \star_4 {\cal{F}} \wedge {\cal{F}} + e^{3A} \frac{X^4}{5} \star_4 g_4 g_4\nn\\[2mm]
& - \frac{g^2}{5} e^{3A} \left(\frac{1}{X^8} - \frac{3}{X^{3}} + 2 X^2 \right)\text{vol}(\text{M}_4)   = 0 \label{eq:scalarEOM}.
\end{align}
The decomposition of Einstein's equations is a bit long and ends up being implied by supersymmetry and the above conditions\footnote{Specifically, \cite{Cariglia:2004qi} prove for any supersymmetric solution of minimal $d=7$ gauged supergravity, where Einstein's equations are ${\cal E}_{\mu\nu}=0$, ${\cal E}_{\mu\nu}\gamma^{\nu}\zeta=0$ is implied by supersymmetry, the Bianchi identities and the equations of motion of the fluxes and scalar. ${\cal V}^{\mu}=\overline{\zeta}\gamma^{\mu}\zeta$ defines a $d=7$ Killing vector. If it is time-like it follows that ${\cal E}_{\mu\nu}=0$ is implied. If it is null, every component of Einstein's equations but ${\cal V}^{\mu}{\cal E}_{\mu\nu}{\cal V}^{\nu}=0$ is implied by ${\cal E}_{\mu\nu}\gamma^{\nu}\zeta=0 $. For the AdS$_3$ solutions we construct  ${\cal V}^{\mu}$ is null, however, given that the condition that fixes ${\cal E}_{\mu\nu}=0$ along the 3 AdS$_3$ directions is common to each of them, we must have that  ${\cal V}^{\mu}{\cal E}_{\mu\nu}{\cal V}^{\nu}=0$ is also implied in this case.} so we will not quote them explicitly.

\subsection{Necessary and sufficient conditions for supersymmetry}\label{eq:nessisarysufficientsusy}
We now turn our attention to deriving a set of necessary and sufficient conditions for AdS$_3$ to preserve supersymmetry. These will be expressed in terms of differential and algebraic constraints on the forms that spann the G-structure that M$_4$ supports.

Our starting is the Killing spinor equations \eqref{eq:susy7dU1} which we need to decompose into their parts among AdS$_3$ and M$_4$.  We make a $7=3+4$ split of the gamma matrices 
\begin{align}
\gamma^{(7)}_{\mu^{(3)}} & = e^{A}\tilde{\sigma}_{\mu^{(3)}} \otimes \hat{\gamma} , ~~~~~~\gamma^{(7)}_a  = \mathbb{I}_2 \otimes \gamma_a ,
\end{align}
where $\tilde{\sigma}_{\underline{\mu}^{(3)}}= (i \sigma_2, \sigma_1, \sigma_3)$ and $\hat{\gamma} = -\gamma_1 \gamma_2 \gamma_3 \gamma_4$. Note that we add a $(7)$ superscript to the $d=7$ gamma matrices here but not earlier to avoid cluttered notation. We like-wise decompose the $d=7$ spinor as
\begin{equation}
\zeta = \zeta^{(3)} \otimes \chi ,
\end{equation}
where $\zeta^{(3)} $ is an AdS$_3$ Killing spinor that obeys
\begin{equation}
\nabla_{\mu^{(3)}}\zeta^{(3)} = \frac{m}{2} \gamma_{\mu^{(3)}} \zeta^{(3)}
\end{equation}
on un-warped AdS$_3$, while $\chi$ is a Dirac spinor on M$_4$. We can use the above to factor out $\zeta^{(3)}$ from \eqref{eq:susy7dU1} and arrive at conditions on M$_4$ alone. The condition that implies the vanishing of the  dilatino variation becomes
\begin{align}
\left( \frac{5}{2} X^{-1} d X + \frac{1}{2} X^2 \left(\hat{\gamma} g_1 + g_4  \right) - \frac{i}{2 \sqrt{2}} X^{-1} {\cal{F}} - \frac{1}{\sqrt{2}} g(X^{-4} - X) \right) \chi=0,\label{eq:dilcod}
\end{align}
while the gravitino condition decomposes into two $d=4$ conditions. First, from the AdS$_3$ directions we get
\begin{align}
&  \left( \frac{m}{2} e^{-A}  + \frac{1}{2} \hat{\gamma} dA + \frac{i}{10 \sqrt{2}} X^{-1} \hat{\gamma} {\cal{F}} - \frac{X^2}{10} g_1 + \frac{3}{20} X^2 \hat{\gamma} g_4 -\frac{g}{5\sqrt{2}} \left( \frac{1}{4} X^{-4} + X \right) \hat{\gamma} \right) \chi = 0.\label{eq:gravcod1}
\end{align}
Second, along the M$_4$ directions we have
\begin{align} 
0 & = \nabla_a \chi + \frac{i}{2} g {\cal{A}}_a \chi - \frac{3 i}{20\sqrt{2}} X^{-1}\gamma_a {\cal{F}} \chi + \frac{i}{4 \sqrt{2}} X^{-1} {\cal{F}} \gamma_a \chi - \frac{1}{10} X^2 \gamma_a g_4 \chi \nn \\[2mm] 
& + \frac{1}{40} X^2 \left( \gamma_a \hat{\gamma} g_1 + 5 \hat{\gamma} g_1 \gamma_a \right) \chi - \frac{g}{5\sqrt{2}} \left(\frac{1}{4} X^{-4} +  X\right)\gamma_a \chi.\label{eq:gravcod2}
\end{align}
Our task now is to construct geometric constraints that imply these 3 conditions.

First off, it is a simple matter to show from the above conditions that\footnote{We could take $\chi_0$ to have any constant norm. The specific value, besides being non-zero, plays no role.}
\beq
d(e^{-A}|\chi|^2)=0~~~~\Rightarrow~~~~ \chi= e^{\frac{A}{2}} \chi_0,~~~~||\chi_0||=1.
\eeq
Further, one can show that the real 1-form component
\beq
K_a=-i \chi^{\dag}\gamma_a\hat\gamma \chi 
\eeq
is such that
\beq
\nabla_{(a}K_{b)}=0,~~~~{\cal L}_K|\chi|^2=0,~~~~{\cal L}_K\chi^{\dag}\hat\gamma\chi=0,
\eeq
which in particular means that either $K=0$, which requires $\chi$ to be chiral, or it is dual to a Killing vector of the entire $d=7$ metric. Before continuing further we find it useful to decompose 
\beq
\chi_0= \cos\left(\frac{\alpha}{2}\right)\eta_++\sin\left(\frac{\alpha}{2}\right)\eta_-,
\eeq
where $\eta_{\pm}$ are chiral spinors with unit norm and $\alpha$ is a point (on M$_4$) dependent phase. Notice this implies
\beq
\chi^{\dag}\hat\gamma\chi= e^{A}\cos\alpha,
\eeq
so that ${\cal L}_K\alpha=0$. As explained in \cite{Apruzzi:2014qva}, such spinors define two SU(2)-structures on M$_4$ spanned by real and holomorphic  2-forms $(J_{\pm},\Omega_{\pm})$ obeying 
\beq
J_{\pm}\wedge \Omega_{\pm}=0,~~~~J_{\pm}\wedge J_{\pm}=\frac{1}{2}\Omega_{\pm}\wedge \overline{\Omega}_{\pm}=2\text{vol}(\text{M}_4),
\eeq
such that
\beq
J_{\pm}=-\frac{i}{2}\eta_{\pm}^{\dag}\gamma_{ab}\eta_{\pm}e^{ab},~~~~\Omega_{\pm}=\mp\frac{i}{2}\eta_{\pm}^{c\dag}\gamma_{ab}\eta_{\pm}e^{ab},
\eeq
where $e^a$ are some vielbein on M$_4$ and $\eta_{\pm}^c=B\eta^*_{\pm}$ for $B$ a hermitian intertwiner such that $BB^*=-\mathbb{I}$ and $B^{-1}\gamma_{a}B=\gamma_a^*$. One can also define a complex vielbein on M$_4$ in terms of orthogonal $(1,0)$ forms $(V,W)$ of norm 2, such that
\beq
ds^2(\text{M}_4)= V\overline{V}+W\overline{W},~~~~\text{vol}(\text{M}_4)=-\frac{1}{4} V\wedge \overline{V}\wedge W\wedge \overline{W}.
\eeq
These one forms are related to the above spinors as
\beq
V_a= \eta_-^{\dag}\gamma_a \eta_+,~~~~~W_a= \eta_-^{c\dag}\gamma_a \eta_+,
\eeq
and to the SU(2)-structure forms as
\beq
J_{\pm}=\pm\frac{i}{2}(V\wedge \overline{V}\pm W\wedge \overline{W}),~~~~\Omega_+=V\wedge W,~~~~\Omega_-=\overline{V}\wedge W,\label{eq:su2forms}
\eeq
such that  $(V,W)$ span the identity-structure common to both SU(2)-structures. 

With what we have collected so far, it is relatively simple to show the equivalence of \eqref{eq:dilcod} and \eqref{eq:gravcod1} to\footnote{Such computations can be greatly expedited by working in a canonical frame in which $(V=e^1+i e^2,W=e^3+i e^4)$ and $\eta_{\pm}$ are constant (though not typically not covariantly constant).} 
\begin{subequations}
\begin{align}
d(e^{A}X)&=\frac{ge^{A}\sin\alpha}{\sqrt{2}2 X^3}\text{Re}V,\label{eq:par1}\\[2mm]
{\cal L}_K A&={\cal L}_K X=0,\label{eq:par2}\\[2mm]
e^{3A}g_1&=\frac{5e^{3A}}{X^3}\cos\alpha dX+ \frac{e^{2A}}{2X^6}\sin\alpha(4m X^4-\sqrt{2}g e^A\cos\alpha)\text{Re}V +\frac{e^{2A}}{\sqrt{2}b_0 c X^3}\iota_{K}{\cal F},\label{eq:par3}\\[2mm]
g_4&=\frac{1}{2 e^{A}X^6}(4mX^4-\sqrt{2}ge^{A}\cos\alpha)\text{vol}(\text{M}_4),\label{eq:par4}\\[2mm]
{\cal F}&=\frac{5\sqrt{2}}{b_0 c e^{A}}\iota_{K}\star_4 dX+ (1-\cos\alpha \star_4)\tilde{{\cal F}} \notag \\[2mm]
&-\frac{1}{e^{A}X^3}\bigg(\cos\alpha( g e^{A}\cos\alpha-2 \sqrt{2}m X^4)-2g e^{A}(1-X^5)\bigg)\text{Re}W\wedge \text{Im}W,\label{eq:par5}
\end{align}
\end{subequations}
where $\tilde{{\cal F}}$ obeys $\text{Im}V\wedge \tilde{{\cal F}}=0$ but is otherwise free. We also now have that
\beq
K= e^{A}\sin\alpha \text{Im}V,
\eeq
with $K=0$ if and only if $\sin\alpha=0$. Given the above and by making use of \eqref{eq:gravcod1}, we can now take the Lie derivative of $\chi$ along $K$. We find
\beq
{\cal L}_K\chi=-i \left(m -\frac{g}{2}(e^{A}\sqrt{2}X\cos\alpha- \iota_K {\cal A})\right) \chi,\label{eq:kosmann1}
\eeq
suggesting that $\chi$ is generically charged under the U(1) isometry of $K$. Indeed, we shall see shortly that the second term on the LHS of this expression is necessarily constant, as this would require.

As it contains a covariant derivative of $\chi$, constructing geometric conditions that are equivalent to  \eqref{eq:gravcod2} is more involved. There is however a, by now well mapped out, path for achieving this\footnote{The methods we make use of in the following are explained at great length in \cite{Tomasiello:2022dwe}, which is a very good resource for the uninitiated.}. We begin by constructing  poly-forms on M$_4$ that are defined in terms of $\chi$ via the bi-linears
\begin{align}
\cancel{\Psi}^1=\chi\otimes \chi^{\dag},~~~~\hat{\cancel{\Psi}}^1=\hat\gamma\chi\otimes \chi^{\dag},\nn\\[2mm]
\cancel{\Psi}^2=\chi\otimes \chi^{c\dag},~~~~\hat{\cancel{\Psi}}^2=\hat\gamma\chi\otimes \chi^{c\dag},
\end{align}
using the Clifford map\footnote{The bilinear \beq
\slashed{\Psi}=\chi_1\otimes \chi^{\dag}_2=\frac{1}{4}\sum_{n=0}^4\frac{1}{n!}\chi^{\dag}\gamma_{a_n...a_1}\chi\gamma^{a_1..a_n}\nn
\eeq
is equivalent to the poly-form $\Psi$ which can be defined by simply replacing $\gamma^{a_1..a_n}$ with $e^{a_1...a_n}$ in the proceeding expression.}.  As poly-forms, these bi-linears are expressed in terms of our complex vielbein as
\begin{align}
\Psi^1&=\frac{ e^{A}}{4}(1+\sin\alpha \text{Re}V)\wedge e^{\frac{1}{2}(\cos\alpha V\wedge \overline{V}+W\wedge \overline{W})},\nn\\[2mm]
\hat\Psi^1&=\frac{ e^{A}}{4}\left(\cos\alpha+\frac{1}{2}V\wedge \overline{ V}+i \sin\alpha\text{Im}V\right)\wedge e^{\frac{1}{2}W\wedge \overline{W}},\nn\\[2mm]
\Psi^2&=-\frac{ie^A}{4}\left(\cos\alpha \text{Re}V+i \text{Im}V-\frac{1}{2}\sin\alpha V\wedge \overline{V}\right)\wedge W,\nn\\[2mm]
\hat\Psi^2&=\frac{ie^A}{4}\left(\sin\alpha-\text{Re}V-\cos\alpha i \text{Im}V\right)\wedge W,\label{eq:bispinors}
\end{align}
where we note that when $\cos\alpha=\pm 1$ these are spanned by the SU(2) structure forms $(J_{\pm},~\Omega_{\pm})$. Through a lengthy computation it is possible to show that \eqref{eq:gravcod2} implies the following differential constraints on the polyform bi-linears:
\begin{subequations}
\begin{align}
&d(e^{-A}\Psi^1_0)=0,\label{eq:Bbps1}\\[2mm]
&d(e^{2A}\Psi^1_2)-e^{2A}\frac{1+4X^5}{2\sqrt{2}X^4}g\Psi^1_3-i \frac{e^{2A}}{\sqrt{2}X}\Psi^1_-\wedge {\cal F}=0,\label{eq:Bbps2}\\[2mm]
&d(e^{3A}\Psi^1_-)=0,\label{eq:Bbps3}\\[2mm]
&d(e^{-A}\hat \Psi^1_0)+\frac{g}{2\sqrt{2}e^A X^4}\cos\alpha\Psi^1_1+\frac{1}{4\sqrt{2}b_0e^{A}X}\iota_{K}{\cal F} =0,\label{eq:Bbps4}\\[2mm]
&d(e^{2A}\hat\Psi^1_2)-2m e^{A}\Psi^1_3=0,\label{eq:Bbps5}\\[2mm]
&d(e^{-2A}\hat\Psi^1_1)+ \frac{g}{e^{2A}\sqrt{2}X^4}\cos\alpha \Psi^1_2+\frac{i }{e^{2A}}\left(\frac{c e^{A}\cos\alpha}{2\sqrt{2}X}{\cal F}- \frac{ X^2}{4 b_0}\iota_K\star_4 g_1\right)=0,\label{eq:Bbps6}\\[2mm]
&d(e^{6A}\hat\Psi^1_3)-6 m e^{5A}\Psi^1_4+\frac{e^{6A}g(1+4X^5)}{\sqrt{2}X^4}\hat\Psi^1_4-\frac{c}{4}e^{7A}X^2\cos\alpha g_4=0,\label{eq:Bbps7}\\[2mm]
&(d+i g{\cal A}\wedge )(e^{3A}\hat\Psi^2_-)- e^{2A}(2m+\frac{ge^{A}}{2\sqrt{2}X^4} \cos\alpha)\Psi^2_++e^{3A}\frac{g(1+4X^5)}{2\sqrt{2}X^4}\hat\Psi^2_+=0,\label{eq:Bbps8}\\[2mm]
&(d+i g{\cal A}\wedge)(e^{7A}\hat\Psi^2_+)+ e^{7A}X^2g_1\wedge \Psi^2_+- e^{6A}(4m+ \frac{3ge^{A}}{2\sqrt{2}X^4}\cos\alpha)\Psi^2_-=0,\label{eq:Bbps9}\\[2mm]
&(d+i g{\cal A}\wedge )(e^{2A}\Psi^2_+)-e^{2A}\frac{g(1+4X^5)}{2\sqrt{2}X^4}\Psi^2_-=0,\label{eq:Bbps10}\\[2mm]
&(d+i g{\cal A}\wedge )(e^{6A}\Psi^2_-)=0\label{eq:Bbps11}.
\end{align}
\end{subequations}
In general this would be not quite enough, as what one needs  are conditions that imply \eqref{eq:gravcod2}. Fortuitously in this case however, with some extra work, one can show that the conditions \eqref{eq:Bbps1}-\eqref{eq:Bbps9} also imply \eqref{eq:gravcod2} (\eqref{eq:Bbps10}-\eqref{eq:Bbps11} being redundant), \textit{i.e.} by expanding the covariant derivative of $\chi$ in terms of complex valued torsion classes as 
\beq
\nabla_{\mu}\chi= (Q^+)_{\mu}^{~a}\gamma_a\eta_-+(Q^-)_{\mu}^{~a}\gamma_a\eta_+,
\eeq
one finds that the components of $(Q^{\pm})_{\mu}^{~a}$ are fixed in the same fashion by either the differential form constraints or by \eqref{eq:gravcod2}. Thus we find that \eqref{eq:par1}-\eqref{eq:par5} and \eqref{eq:Bbps1}-\eqref{eq:Bbps9} are necessary and sufficient conditions for supersymmetric AdS$_3$ solutions.

It turns out the above conditions are consistent with two classes of solutions, which we derive in the next sections. Before let us make some further comments about the Killing vector $K$. By making use of \eqref{eq:Bbps4} one can now show that
\beq
\iota_{K}{\cal F}= - \sqrt{2}d(e^{A}X\cos\alpha),
\eeq
which, when $\sin\alpha\neq 0$, fixes the portion of ${\cal F}$ parallel to $K$ such that we must have
\beq
\iota_{K}{\cal A}= (\sqrt{2}e^{A}X\cos\alpha+ \frac{2a_0}{g})
\eeq
for $a_0$ some constant. If we return to \eqref{eq:kosmann1} we now find that
\beq
{\cal L}_K\chi=-i (m +a_0) \chi,
\eeq
making it clear that $\chi$ is charged under $K$ unless $a_0=-m$. Let us stress that this computation only holds weight when $\sin\alpha\neq 0$, a condition that the first class we consider violates.

\subsection{Kahler-Einstein class}\label{eq:sec:KEclass}
In this section we consider the class for which $\sin\alpha=0$. We can without loss of generality take $\alpha=0$ such that $\chi$ has positive chirality. Fixing $\alpha=0$ means that $K=0$, so supersymmetry does not impose a Killing vector on M$_4$ and it is possible to show that \eqref{eq:par1}-\eqref{eq:par5} and \eqref{eq:Bbps1}-\eqref{eq:Bbps11} reduce to
\begin{align}
&e^{A}=\frac{\sqrt{2}m}{g X},\\[2mm]
&d(e^{3A}J_+)=0,~~~~d(e^{3A}\Omega_+)+i g {\cal A}\wedge(e^{3A}\Omega_+)=0,\label{eq:case1definM4}\\[2mm]
&e^{3A}g_1=\frac{10\sqrt{2}m^3}{X^6}dX,~~~~g_4= -g \frac{1-2 X^5}{\sqrt{2}X^6}\text{vol}(\text{M}_4),~~~~{\cal F}= \frac{g^4}{4\sqrt{2}m^3}e^{3A}J_+.
\end{align} 
The condition \eqref{eq:case1definM4} implies that M$_4$ is conformally a Kahler-Einstein manifold. Specifically, we have that
\beq
ds^2(\text{M}_4)= e^{-3A}ds^2(\text{KE}_4),~~~~ \hat{J}= e^{3A}J_+,~~~~\hat{\Omega}= e^{3A}\Omega_+,
\eeq
where $(\hat{J},\hat{\Omega})$ define the Kahler and holomorphic 2-forms on the unwarped Kahler-Einstein manifold. The remaining conditions just fix the forms and functions of our ansatz in terms of the SU(2)-structure on this manifold and $X$, which can generically have support on KE$_4$. At this point the necessary conditions for supersymmetry are solved.

In summary we find the following class of solutions
\begin{align}
ds^2&= e^{2A}ds^2(\text{AdS}_3)+ e^{-3A}ds^2(\text{KE}_4),~~~~e^{A}=\frac{\sqrt{2}m}{g X},\\[2mm]
{\cal F}&=\frac{g^4}{4\sqrt{2}m^3}\hat{J},~~~e^{3A}g_1=\frac{10\sqrt{2}m^3}{g^3X^6}dX,~~~~g_4= \frac{g}{\sqrt{2}X^6}(2 X^5-1)e^{-6A}\text{vol}(\text{KE}_4),
\end{align}
where $\hat{J}$ is the Kahler form on KE$_4$.
Supersymmetry demands that
\beq
d\hat J=0,~~~~~d\hat\Omega+i g {\cal A}\wedge \hat{\Omega}=0.\label{eq:susycond}
\eeq
The flux EOM \eqref{fluxeom1} and \eqref{fluxeom2} are implied while \eqref{fluxeom3} reduces to
\beq
10 d(X \star_4dX)+ \frac{g^8}{4 m^6}(X^5-\frac{3}{4})\text{vol}(\text{KE}_4)=0.
\eeq
When this holds \eqref{eq:scalarEOM} is implied, so all the field equations then follow. Solutions in this class preserve 4 supercharges, or ${\cal N}=(2,0)$ from the AdS$_3$ perspective.

Note that the dependence of $g$ in $e^{A}$ coupled with how $g$ appears in the metric requires $g>0$. Then \eqref{eq:susycond} implies first that KE$_4$ is indeed a Kahler manifold. Then, given that $d{\cal A}= \frac{g^4}{4\sqrt{2}m^3}\hat{J}$, that it is also Einstein. Finally, the lack of a relative sign between the terms in the second of \eqref{eq:susycond} tells us that KE$_4$ has negative curvature. A specific case of this class, in which $X$ is constant, was found before in \cite{Gauntlett:2000ng}, where it appeared as the IR fixed point of a flow from AdS$_7$.

\subsection{U(1) fibration over $\Sigma\times \mathbb{R}$}\label{sec:class2}
In this section we consider the class that follows from assuming that $\sin\alpha\neq 0$, for which $K$ defines a Killing vector on M$_4$. This class appears to be more rich and contains all the explicit solutions we construct later in the paper.

It is possible to show that the necessary and sufficient conditions for supersymmetry in this case reduce to the following differential conditions:
\begin{subequations}
\begin{align}
&{\cal L}_KA={\cal L}_K X={\cal L}_K \alpha=0,\label{eq:case2bps0}\\[2mm]
&d(e^{A}X)-\frac{g e^A}{2\sqrt{2}X^3}\sin\alpha\text{Re}V=0,\label{eq:case2bps1}\\[2mm]
&d(e^{4A}\sin\alpha W)+i ge^{3A}\left(e^{A}\sin\alpha {\cal A}+\left(\frac{2m}{g}-\sqrt{2}e^{A}X\cos\alpha\right)\text{ImV}\right)\wedge W\nn\\[2mm]
&+e^{3A}\frac{8m X^4 \cos\alpha-\sqrt{2}e^{A}g(\sin^2{\alpha}+4X^5)}{4 X^4} \text{Re}V\wedge W=0,\label{eq:case2bps2}\\[2mm]
&\sqrt{2}b_0 c  d(e^{A}X\cos\alpha)+ \iota_{K}{\cal F}=0,\label{eq:case2bps3}\\[2mm]
&d(e^{3A}(\cos\alpha V\wedge \overline{V}+W\wedge \overline{W}))-i\frac{\sqrt{2} e^{3A}\sin\alpha}{X}\text{Re}V\wedge\left({\cal F}-\frac{i g(1+4 X^5)}{4 X^3} W\wedge \overline{W}\right)=0,\\[2mm]
& d(e^{-A} \sin\alpha \text{Im}V)+\frac{\sqrt{2}}{e^{A}X}\cos\alpha {\cal F}- \frac{X^2}{e^{2A} b_0 c}\iota_K\star_{4} g_1- \frac{i g\cos\alpha}{2 \sqrt{2}e^{A}X^4}\left(\cos\alpha V\wedge \overline{V}+W\wedge \overline{W}\right)=0, \\[2mm]
{\cal F}&=\frac{5\sqrt{2}}{b_0 c e^{A}}\iota_{K}\star_4 dX+ (1-\cos\alpha \star_4)\tilde{{\cal F}} \notag \\[2mm]
&-\frac{1}{e^{A}X^3}\bigg(\cos\alpha( g e^{A}\cos\alpha-2 \sqrt{2}m X^4)-2g e^{A}(1-X^5)\bigg)\text{Re}W\wedge \text{Im}W,
\end{align} 
\end{subequations}
and the following flux definitions
\begin{align}
e^{3A}g_1&=d\left(\frac{2\sqrt{2}m(e^{A}X)^2}{g}-e^{3A}X^{-2}\cos\alpha\right),\nn\\[2mm]
g_4&=\frac{4 m X^4-\sqrt{2}g e^{A}\cos\alpha}{2 e^{A}X^6}\text{vol}(\text{M}_4),\label{eqs:fluxes}
\end{align}
given that $\sin\alpha\neq 0$ and $K$ is dual to a Killing of M$_4$ vector.

We begin by solving \eqref{eq:case2bps1} in terms of a local coordinate $y$ and function $p(y)$ as
\beq
e^{A}X=p,~~~~ \text{Re}V=\frac{2\sqrt{2}p^3 p'}{g e^{4A}\sin\alpha}dy.
\eeq
Note that, beyond not being constant, $p$ is arbitrary and essentially parametrises diffeomorphism invariance in $y$.
Then, given that $K^a\partial_a$ is a Killing vector and $K$ is parallel to $\text{Im}V$, we are free to parametrise
\beq
\text{Im}V=\frac{1}{c_0}e^{A}\sin\alpha D\psi,~~~~ D\psi=d\psi+{\cal B},
\eeq 
where ${\cal B}$ has no leg in $dy$ and is independent of $\psi$. The internal space decomposes as
\beq
ds^2(\text{M}_4)=\frac{8 p^6 (p')^2}{g^2 e^{8A}\sin^2\alpha}dy^2+ g_{ij}(y,z_i)dz_i dz_j+ e^{2A}\sin^2\alpha \frac{1}{c_0^2}D\psi^2
\eeq
where $W\overline{W}= g_{ij}(y,z_i)dz_i dz_j$.  We now have from \eqref{eq:case2bps0} that $(e^A,X,\alpha)$ are independent of $\psi$ and from \eqref {eq:case2bps3} that
\beq
{\cal F}=\frac{\sqrt{2}}{c_0} d(e^{A}X\cos\alpha D\psi)+\hat{{\cal F}},
\eeq
where ${\cal \hat{F}}$ is orthogonal to $D\psi$ but is otherwise to be determined. For $d{\cal F}=0$ to hold there should locally exist a potential $\hat {\cal A}$ such that
\beq
d\hat {\cal A} =\hat{{\cal F}}.
\eeq
As such we can decompose the gauge field as
\beq
{\cal A}= \frac{\sqrt{2}}{c_0} e^{A}X\cos\alpha D\psi+\hat{\cal A},
\eeq
where we are free to choose gauge in which $\hat{\cal A}$ has no leg in $dy$. We elect to do this.  At this point it is useful to decompose the exterior derivative as
\beq
d= d\psi\wedge \partial_{\psi}+dy \wedge \partial_y+d_2.
\eeq
 Give this, we find that \eqref{eq:case2bps2}  decomposes as
\begin{align}
&\partial_{\psi}(e^{4A}\sin\alpha W)+i \frac{2m}{c_0}  (e^{4A}\sin\alpha W)=0,\nn\\[2mm]
&\partial_{y}(e^{4A}\sin\alpha W)- f (e^{4A}\sin\alpha W)=0, \nn\\[2mm]
&d_2(e^{4A}\sin\alpha W)+ i \rho\wedge (e^{4A}\sin\alpha W)=0,\nn\\[2mm]
&\hat\rho=(g \hat{\cal A}+ \frac{2m}{c_0}{\cal B}),\nn\\[2mm]
&f=-\frac{\frac{ 4 \sqrt{2}m}{g}X^4\cos\alpha+e^{A}(\cos^2\alpha-(1+4 X^5))}{e^{5A}X^4 \sin^2\alpha}p^3 p'.
\end{align}
These are solved by taking
\beq
e^{4A}\sin\alpha W= e^{-\frac{2m i}{c_0}\psi}e^{\mu}(dz_1+i dz_2),
\eeq
where $\mu=\mu(y,z_i)$ such that
\beq
\partial_{y}\mu=f,~~~~ \hat\rho=\partial_{z_1}\mu dz_2-\partial_{z_2}\mu dz_1.
\eeq
At this point it is easy to extract all the supersymmetry constraints. We have
\begin{align}
{\cal A}&= \frac{\sqrt{2} p}{c_0}\cos\alpha D\psi+\hat{\cal A},~~~~{\cal B}=\frac{c_0}{2m}\left(\partial_{z_1}\mu dz_2-\partial_{z_2}\mu dz_1-g \hat{\cal A}\right), \label{eqs:guage1}
\end{align}
and we can decompose
\beq
\hat{\cal A}= a_1(y,z_i)dz_1+ a_2(y,z_i)dz_2.
\eeq
The remaining supersymmetry constraints then impose that $(a_1,a_2)$ are constrained as
\begin{align}
\partial_ya_i&=-\frac{4p'}{g}\epsilon_{ij}\partial_{z_j}\left(\frac{p^4}{e^{5A}\sin^2\alpha}\right),\nn\\[2mm]
\partial_{z_2}a_1-\partial_{z_1}a_2&=\frac{g}{2pp'}\left(\partial_{y}\left(\frac{e^{2\mu}}{e^{5A}p\sin^2\alpha}\right)-\frac{e^{2\mu}4p^3p'}{e^{10A}\sin^2\alpha}\right),\label{eqs:guage2}
\end{align}
which essentially define the components of $\hat{\cal F}$. It also follows that $\mu$ must obey the conditions
\begin{subequations}
\begin{align}
&\partial_y\log(e^{\mu}p^{-1})-\frac{4 p^3p'(gp-\sqrt{2}m \cos\alpha)}{g e^{5A}\sin^2\alpha}=0,\label{eq:bps1}\\[2mm]
&\partial_{z_i}^2\mu+\frac{g}{2p p'}\bigg[\partial_{y}\left(\frac{e^{2\mu}(gp-\sqrt{2}m \cos\alpha)}{e^{5A}\sin^2\alpha p^2}\right)-\frac{2 p' e^{2\mu}(\sqrt{2}gm e^{5A}\cos\alpha-4m^2 p^4+ 2g^2 p^6)}{g e^{10A}\sin^2\alpha p^3}\bigg]=0 \label{eq:bps2}.
\end{align}
\end{subequations}
At this point supersymmetry is solved.

To have a solution one still needs to satisfy the equations of motion of the theory and the Bianchi identities of the fluxes. The only Bianchi identity that is not implied by supersymmetry is  $d{\cal F}=0$ which, amount to imposing this for $\hat{\cal F}$, leads to
\begin{align}
4\partial_{z_i}^2\left(\frac{p^4p'}{g e^{5A}\sin^2\alpha}\right)+ \partial_y\bigg(\frac{g}{2p p'}\partial_{y}\left(\frac{e^{2\mu}}{e^{5A}p\sin^2\alpha}\right)-\frac{2 g e^{2\mu} p^2}{e^{10A}\sin^2\alpha}\bigg)=0.\label{eq:bi1}
\end{align}
It turns out that when this holds the remaining equations of motion are implied. Thus we have found a local class of solutions for which the metric takes the form
\beq
ds^2(\text{M}_4)=\frac{1}{e^{8A}\sin^2\alpha}\bigg[\frac{8 p^6 (p')^2}{g^2 }dy^2+ e^{2\mu}(dz_i)^2\bigg]+ e^{2A}\sin^2\alpha \frac{1}{c_0^2}D\psi^2,
\eeq
the fluxes $(e^{3A}g_1,g_4)$ are fixed by \eqref{eqs:fluxes} and gauge field by \eqref{eqs:guage1}-\eqref{eqs:guage2}. Each solution to \eqref{eq:bps1}, \eqref{eq:bps2} and \eqref{eq:bi1} defines a solution in the class.

Being that it requires solving 3 independent PDEs to define a supersymmetric solution, this class appears to be rather complicated. With that said, such complication tends to be an indication that a class contains quite a wide variety of solutions.  To try and make headway in finding some of these, we shall make a simplifying assumption in the next section, namely that $\Sigma$ has constant curvature. 

\section{AdS$_3$ solutions containing constant curvature Riemann surfaces}\label{sec:CC}
The equations defining the class in the previous section are rather complicated and non-linear. As such, let us make the simplifying assumption that  $(z_1,z_2)$ span a Riemann surface of constant curvature. This can be achieved by imposing a separation of variables ansatz
\beq
e^{\mu}= e^{\Delta(z_i)}P(y),
\eeq
where we assume that
\beq
\partial_{z_i}^2\Delta+ \kappa e^{2\Delta}=0,
\eeq
which yields\footnote{For $\kappa \neq 0$ really $\kappa=\pm |\kappa|$ is what distinguish S$^2$ and $\mathbb{H}^2$, but $\kappa=\pm 1$ gives a unit radius maximally symmetric space. More general values of $\kappa \neq 0$ do not lead to more general solutions as $|\kappa|$ can always be rescaled out of the ansatz.}
\beq
 ds^2(\Sigma)=e^{2\Delta}(dz_1^2+dz_2^2)=\left\{\begin{array}{l}ds^2(\text{S}^2)~~~~\kappa=1\\[2mm]
ds^2(\mathbb{T}^2)~~~~\kappa=0\\[2mm]
ds^2(\mathbb{H}^2)~~~~\kappa=-1.\end{array}\right. 
\eeq
Each of which makes $\Sigma$ compact, provided that we take a discrete quotient in the case of $\mathbb{H}^2$. It is a simple matter to check that  \eqref{eq:bps1}, \eqref{eq:bps2} and \eqref{eq:bi1} then impose that $e^{A},\alpha$ are functions of $y$ only. It is a simple matter to check that imposing that $e^{A},\alpha$ are functions of $y$ only, allows \eqref{eq:bps1}, \eqref{eq:bps2} and \eqref{eq:bi1} to be written as
\begin{subequations}
\begin{align}
&\frac{P^2}{p e^{5A}\sin^2\alpha}\left(g-\frac{\sqrt{2}m \cos\alpha}{p}\right)-\frac{g}{8 p' p^3}\partial_y\left(\frac{P^2}{p^2}\right)=0,\label{eq:CC1}\\[2mm]
&\partial_{y}\left(\frac{P^2}{p e^{5A}\sin^2\alpha}-\frac{c_1 p^2}{g^2}\right)- \frac{4 P^2 p^3  p'}{e^{10A}\sin^2\alpha}=0,\label{eq:CC2}\\[2mm]
&\partial_{y}\left(\frac{\sqrt{2}m P^2\cos\alpha}{p^2e^{5A}\sin^2\alpha}-\frac{c_1-\kappa}{g}p^2\right)-\frac{2 p' p^3 P^2}{g e^{10A}\sin^2\alpha}\left(\frac{4m^2}{p^2}-\frac{\sqrt{2}gm e^{5A}\cos\alpha}{p^6}\right)=0,\label{eq:CC3}
\end{align}
\end{subequations}
where $c_1$ is an integration constant. We now find it convenient to use diffeomorphism invariance to fix
\beq
p= \frac{m}{g}\sqrt{y},
\eeq
and to introduce $h$ and $G$, two arbitrary functions of $y$ such that
\beq
h=\frac{g^5 P^2}{m^3 \sqrt{y}e^{5A}\sin^2\alpha}- c_1 y,~~~G=\frac{\sqrt{2}g^2 P^2 \cos\alpha}{my e^{5A}\sin^2\alpha}- \frac{m^2 y(c_1-\kappa)}{g^3},
\eeq
where it follows that $h$ cannot be constant. Using these to eliminate $(\alpha, P)$ we  find that \eqref{eq:CC2}-\eqref{eq:CC3} are solved as
\beq
e^{5A}= \frac{ 2m^5 y^{\frac{3}{2}}(c_1 y+h)}{g^5 h'},~~~~~G= m^2 \frac{c_2+4 h-(c_1-\kappa)y^2}{2g^3 y},
\eeq
where $c_2$ is a second integration constant. Note that as $e^{A} \sim m$, which is the inverse radius of AdS$_3$, we effectively have
\beq
m=1,
\eeq
which we will simply assume going forward. We are now only left with \eqref{eq:CC1} to solve, which defines the solutions, and we can extract the form of the bosonic fields in terms of $h$ and its derivatives.

In summary the $d=7$ bosonic fields now take the form
\begin{align}
g^2 ds^2&=y^{\frac{3}{5}}\left(\frac{\Lambda}{h'}\right)^{\frac{2}{5}} \bigg[ds^2(\text{AdS}_3)+ \frac{2 (h')^2 dy^2 }{y \Lambda^2(1- \frac{\Xi^2}{2 y \Lambda^2})}+\frac{(1- \frac{\Xi^2}{2 y \Lambda^2}) n^2}{4}D\psi^2+ \frac{h'}{2 y}ds^2(\Sigma)\bigg],\nn\\[2mm]
g{\cal A}&= c_1\rho -\frac{n \Xi}{2 \Lambda}D\psi,~~~~ D\psi=d\psi+\frac{c_1-\kappa}{n}\rho,~~~~X= \left(\frac{y h'}{\Lambda}\right)^{\frac{1}{5}},\nn\\[2mm]
g^3 {\cal F}_4&= \frac{1}{\sqrt{2}}\text{vol}(\text{AdS}_3)\wedge d\left(4y-\frac{\Xi}{h'}\right)+\frac{n(4y h'-\Xi)}{4\sqrt{2} y^2} D\psi\wedge dy\wedge \text{vol}(\Sigma),\label{eq:genconstsol}
\end{align}
where AdS$_3$ has unit radius (\textit{i.e.} $m=1$). We have fixed/defined
\beq
c_0=-\frac{2}{n},~~~~\Lambda=2 (c_1 y+h),~~~~\Xi= c_2+(c_1-\kappa)y^2+4 h,
\eeq
and the connection $\rho$ is such that $d\rho= -\text{vol}(\Sigma)$ and takes the form
\beq
\rho= \left\{\begin{array}{l} \frac{1}{2}(z_2 dz_1-z_1 dz_2),~~~~~\kappa=0,~~~\Delta=0\\[2mm]
  \frac{1}{\kappa}\epsilon_{ij} \partial_{z_i}\Delta dz_j,~~~~~~~~~~~~\kappa=\pm 1.\end{array}\right.
\eeq
Solutions are defined by the single second order ODE
\beq
\left(\Xi^2-2y \Lambda^2\right)h''+4\left(y \kappa \Xi+\frac{\Lambda^2}{2}- c_1 y(c_2+y(c_1(y-4)-\kappa y))\right)h'=0.\label{eq:CCPDE}
\eeq
\subsection{Flux quantisation and holographic central charge}
For the constant curvature class of this section we can compute the holographic central charge through the formula
\begin{equation}\label{eq:cc2d}
c_{\text{hol}} = \frac{3}{2^5 \pi^7} \int_{\text{M}_8} e^A \text{vol}(\text{M}_8).
\end{equation}
This formula requires us to uplift the 7d background to 11d and then integrate over the internal directions. We do this by using the uplift of \eqref{eq:11dmet}. Flux quantisation in $d=11$ requires that the 4-form $G$ obeys
\begin{equation}\label{eq:fluxquantG11d}
\frac{1}{(2 \pi)^3} \int_{\text{S}^4} G = N,
\end{equation}
where $N$ is the number of M5 branes wrapping M$_4$. This fixes
\begin{equation} 
L^3 = \frac{ \tilde{k} N \pi }{2 \sqrt{2}}.\label{eq:Lfixed}
\end{equation}
As we seek solutions containing a compact M$_4$, it is also possible to define a charge at the poles of the 4-sphere in \eqref{eq:11dmet} where $\sin\beta=\pm 1$. This leads to
\begin{equation}
\frac{1}{(2 \pi)^3} \int_{\text{M}_4^{\pm}} G= \pm \frac{\tilde{k} N n }{2^7 \pi^2 } \int_{\text{M}_4^{\pm}}\frac{ \left(\Xi- 4 y h'  \right)}{y^2} d\psi \wedge dy \wedge  \text{vol}(\Sigma) , \label{eq:4formoversigma4}
\end{equation}
but unlike the previous charge, we require a solution to $h$ and global details about M$_4$ to perform the integral. 
We then find that the details of the embedding manifold S$^4$ drop out of \eqref{eq:cc2d}, leading to
\begin{equation}
c_{\text{hol}} =  \frac{N^3\tilde{k}^2}{64 \pi^2}\int_{\text{M}_4} dy\wedge d \psi\wedge \text{vol}(\Sigma)  n h', \label{eq:holcent}
\end{equation}
where again one needs a solution to $h$ in hand and global details about M$_4$ to perform this integral.

\subsection{U(1) R-symmetry Killing vector}
One can compute which Killing vector is dual to the  U(1) R-symmetry of the constant curvature Riemann surface class of solutions through  vector bi-linears formed from the spinor the class of solutions supports, we maintain the conventions of section \ref{eq:nessisarysufficientsusy} with $m=1$. To this end we must solve the Killing spinor equation on AdS$_3$
\beq
\nabla_{\mu^{(3)}}\zeta^{(3)}= \frac{1}{2}\tilde{\sigma}_{\mu^{(3)}}\zeta^{(3)}.
\eeq
To proceed we parametrise the metric and vielbein as
\beq
ds^2(\text{AdS}_3)= e^{2r}(-dt^2+d\hat{x}^2)+dr^2,~~~~~~ e^{a}=(e^{r} dt,~ e^{r} d\hat{x},~ dr)^a.
\eeq
With respect to this choice we find that $\zeta^{(3)}$ decomposes as
\beq
\zeta^{(3)}=b_1 \zeta^{(3)}_1+b_2 \zeta^{(3)}_2,~~~~\zeta^{(3)}_1=e^{\frac{r}{2}}\left(\begin{array}{c}1\\0\end{array}\right),~~~~\zeta^{(3)}_2=\left(e^{-\frac{r}{2}}+ e^{\frac{r}{2}}( t \tilde{\sigma}_1+ \hat{x} \tilde{\sigma}_2\right)\left(\begin{array}{c}0\\1\end{array}\right),
\eeq
where $b_{1,2}$ are constants and $\zeta^{(3)}_{1,2}$ are respectively the Poincar\'e and conformal supercharges on AdS$_3$, which are independent.

We can now use these independent supercharges on AdS$_3$ to define two spinors on the entire $d=7$ manifold, namely 
\beq
\zeta_a=\zeta^{(3)}_a\otimes \chi,~~~a=1,2,
\eeq
where $\chi$ is defined as in section \ref{eq:nessisarysufficientsusy} but refined to the constant curvature Riemann surface ansatz. We can define 3 non-zero vector bi-linears out of these spinors:
\begin{align}
\overline{\zeta}_1(\gamma^{(7)})^{\mu}\zeta_1&= -(\partial_t+\partial_{\hat{x}}),\nn\\[2mm]
\overline{\zeta}_2(\gamma^{(7)})^{\mu}\zeta_2&= -(e^{-2r}+ (t+\hat{x})^2)\partial_t+(e^{-2r}- (t+\hat{x})^2)\partial_{\hat{x}}+2(t+\hat{x})\partial_{r}\\[2mm]
\overline{\zeta}_1(\gamma^{(7)})^{\mu}\zeta_2&= - (t+\hat{x})\partial_{t}- (t+\hat{x})\partial_{\hat{x}}- i \frac{2}{n}\partial_{\psi}.\nn
\end{align}
The real parts of each of these expressions give the 3 independent Killing vectors on AdS$_3$ spanning the SL(2)$_+$ subgroup of SO(2,2)$\equiv$SL(2)$_+\times $SL(2)$_-$. In fact, what appears are precisely the Killing vectors that generate the $d=2$, ${\cal N}=(2,0)$ superconformal algebra if we identify the U(1) R-symmetry Killing vector as
\beq \label{eq:Rsymmsugra}
R=-\frac{2}{n}\partial_{\psi},
\eeq
which in particular only depends on the U(1) fibre isometry in all cases, though the specifics of $n$ will depend on the solution to \eqref{eq:CCPDE} and the global details of that solution. 
\subsection{Polynomial solutions}\label{sec:polysols}
While in  section \ref{sec:CC} we arrived at a more manageable system than the general one, the ODE \eqref{eq:CCPDE} is still highly non-linear and so finding a general solution seems out of the question. Thus we instead seek polynomial solutions. We are in fact able to find 3 of them:
\begin{subequations}
\begin{align}
h&=\frac{c_2(c_1-\kappa)}{4\kappa}- \frac{c_1^2}{\kappa}y+\frac{(c_1-\kappa)^2}{4\kappa}y^2~||~\kappa \neq 0,\label{eq:CCsol1}\\[2mm]
h&= b_1-4 b_2 y+ b_2 y^2~||~c_1=c_2=\kappa=0,\label{eq:CCsol2}\\[2mm]
h&=-\frac{c_2(c_1-\kappa)}{4(2c_1-\kappa)}-\frac{3c_1^2}{2c_1-\kappa}y+ \frac{3(c_1-\kappa)^2}{4(2c_1-\kappa)}y^2\label{eq:CCsol3},
\end{align}
\end{subequations}
where $(b_1,b_2)$ are arbitrary constants. Let us comment on these local solutions in turn:

Though \eqref{eq:CCPDE} does not demand this, it is possible to show that metric positivity for the local class that \eqref{eq:CCsol1} defines requires $\kappa>0$, so it is only valid for $\Sigma=$S$^2$. Further this class has ${\cal F}=d{\cal A}=0$. We show in appendix \ref{sec:recovingtheitalians} that this class contains 2 known solutions that take the local form of AdS$_3\times$S$^3$ foliated over an interval whose uplift to $d=11$ preserves the small and large ${\cal N}=(4,4)$ superconformal algebras, as explored in \cite{Conti:2024rwd,Conti:2024rqy}.

The local solution that \eqref{eq:CCsol2} gives rise to is only compatible with ${\Sigma}=\mathbb{T}^2$. However, the interval spanned by $y$ in this case does not appear to be bounded by physical behaviour, so we will not discuss this solution further.

The most interesting of the 3 local solutions is \eqref{eq:CCsol3}, for which one can quickly establish that $d{\cal A}\neq 0$. The solution admits many global completions depending on how $(c_1,c_2,\kappa)$ are fixed. Indeed the analysis of the new global solutions it contains, for which M$_4$ is compact, and their CFT duals constitute  the rest of this paper. It contains two further known solutions we recover in appendix \eqref{eq:knownsols2}, for which M$_4$ is locally a product of $\mathbb{H}^2$ and a spindle. Before moving onto this analysis in the next sections  allow us to pause to make a few comments which might be instructive for follow up work:

In the two solutions that follow from \eqref{eq:CCsol1} the $y$ interval is only bounded from below, however the resulting geometries can still be interpreted in an AdS/CFT fashion as they asymptote to AdS$_7$  \cite{Conti:2024rwd,Conti:2024rqy}. While we do not consider this possibility in this work, it may be fruitful to consider what asymptotically AdS$_7$ solutions \eqref{eq:CCsol3} contains. Indeed, provided that $c_1 \neq \kappa$, we find that the  metric tends to
\begin{equation}
ds^2 \to \frac{8}{g^2} \left( \frac{y}{8} ds^2(\text{AdS}_3) + \frac{1}{4 y^2} dy^2 + \frac{y}{32} n^2 D \psi^2 + \frac{3 y (c_1 - \kappa)^2}{32 (2 c_1 - \kappa)} ds^2(\Sigma) \right),
\end{equation}
as  $y \to \infty$. This is locally AdS$_7$ with radius $L = \frac{g}{2 \sqrt{2}}$,  \textit{i.e.} we have at this point that
\begin{equation}
R_{\mu \nu \rho \sigma} = - \frac{g^2}{8} (g_{\mu \rho} g_{\nu \sigma} - g_{\mu \sigma} g_{\nu \rho}).
\end{equation}
Such solutions may enjoy a holographic description in terms of, for instance, co-dimension 4 conformal defects - however that remains to be seen. A second point is that we are strictly only considering solutions to \eqref{eq:CCsol3} that do not give rise to curvature singularities, \textit{i.e.} we are ignoring scenarios such as the $y$ interval being bounded by a point where the AdS$_3$ warp either vanishes or blows up. But there are such singular solutions which do give rise to well defined solutions, a prominent example is branes that wrap a topological disc as in \cite{Bah:2021hei}. It may be worth exploring whether M$_4$ can yield a $d=4$ analogue of this. 

We now proceed with our analysis of the distinct global solutions, with compact internal spaces and no curvature singularities, that \eqref{eq:CCsol3} give rise to.

\section{M5 branes wrapping $\mathbb{WCP}^2_{[k,k,\ell]}$}\label{eq:M5braneswrappingWCP2}
In this section we construct our first globally well defined supersymmetric solution that follows from the polynomial class of \eqref{eq:CCsol3}. The uplift of this solution to $d=11$ yields a metric for M5 branes compactified on a topological $\mathbb{CP}^2$ with two orbifold fixed points which we identify as the weighted projective space $\mathbb{WCP}^{2}_{[k,k,\ell]}$.
    
To construct this solution we begin by taking $h$ as in \eqref{eq:CCsol3} and tuning 
\beq
c_2= \frac{4 c_1^3(c_1-2\kappa)}{(c_1-\kappa)^3},
\eeq
then substitute the result  into \eqref{eq:genconstsol}. With this tuning we see that the numerator and denominator of the AdS$_3$  warp factor
\beq
e^{2A}=\frac{y^{\frac{3}{5}}}{g^2}\left(\frac{2 (c_1 y+h)}{h'}\right)^{\frac{2}{5}}
\eeq
share a common factor, which they do not for generic $c_2$.  This allows for the $(D\psi^2,ds^2(\Sigma))$ dependent parts of the metric to both vanish at a point where $e^{2A}$ remains finite -- if we take $\Sigma =\text{S}^2$, this point is locally the origin of  $\mathbb{R}^4$. We find it helpful to redefine 
\beq
c_1=\frac{c-1}{2c},~~~~\kappa=1,~~~~ n=\frac{c+1}{c\ell},~~~~ y=2 \left(\frac{c-1}{c+1}\right)^2 x,
\eeq
upon which the local solution then takes the form
\begin{align}
ds^2&= \frac{2}{g^2} \left( \frac{ (1-c)^4 x^{\frac{3}{2}}H}{3} \right)^{\frac{2}{5}} \bigg[\frac{1}{(c+1)^2} ds^2(\text{AdS}_3)+ds^2(\Sigma_4)\bigg],~~~~X^5=\frac{3(1-c)x}{H},\nn\\[2mm]
ds^2(\Sigma_4)&=\frac{9}{(1-x)P}dx^2+(1-x)\bigg(\frac{ P}{4x c^2\ell^2 H^2}D\psi^2+\frac{3}{16cx}ds^2(\text{S}^2)\bigg),\nn\\[2mm]
g{\cal A}&=\frac{2(3c+1)(1-c)(1-x)}{c\ell H}D\psi-\frac{1-c}{c\ell}d\psi,~~~~D\psi= d\psi-\frac{\ell}{2}\rho,~~~~d\rho=-\text{vol}(\text{S}^2),\nn\\[2mm]
{\cal F}_4&=\frac{(3c+1)(c-1)^2}{g^3}\bigg[\frac{2\sqrt{2}}{3(c+1)^3}\text{vol}(\text{AdS}_3)\wedge dx+ \frac{(x^2-1)}{8 \sqrt{2}c^2 \ell x^2}dx\wedge d\psi\wedge \text{vol}(\text{S}^2)\bigg],
\end{align}
where we define the functions
\beq
P=16(3c+1)x-H^2,~~~~H=3c+1+3(1-c)x.
\eeq
Provided that $0<c<1$, $P$ has two real roots which lie on either side of $x=1$,  the warp factors of the metric are positive between the smaller root, $x=x_0$, and $x=1$. As such, the $x$ interval is bounded as $0<x_0\leq x\leq 1$ for 
\beq
x_0=\frac{(3c+1)(2-\sqrt{3c+1})^2}{9(c-1)^2}.
\eeq

As the overall warping of the above metric has no zero or infinity across the entire $x$ interval, the topology of the internal space is determined by $\Sigma_4$ alone. We can study the behaviour as $x\to 1$ by defining $x=1-z^2$ and expanding in small $z$. We find
\beq
ds^2(\Sigma_4)\to \frac{3}{4c}\bigg(dz^2+z^2\left(\frac{1}{4}ds^2(\text{S}^2)+ \frac{1}{\ell^2}D\psi^2\right)\bigg),
\eeq
which is an $\mathbb{R}^4/\mathbb{Z}_{\ell}$ orbifold fixed point provided that $\psi$ has period $2\pi$ and  $\ell\in \mathbb{Z}$. Conversely, as $x\to x_0$ we find
\beq
ds^2(\Sigma_4)\to \frac{36}{(1-x_0)P'(x_0)}\left(dz^2+ \left(\frac{(x_0-1)P'(x_0)}{12 c \ell \sqrt{x_0}H(x_0)}\right)^2z^2 D\psi^2\right)+\frac{3(c+1)^2(1-x_0)}{16 c x_0} ds^2(\text{S}^2),
\eeq
where $x=z^2+x_0$. This time the S$^2$ factor is a positive constant while only the  $D\psi^2$ direction is vanishing. If we ignore the fibration the $(x,D\psi)$ directions are behaving like an $\mathbb{R}^2/\mathbb{Z}_k$ orbifold fixed point if we tune
\beq
c= \frac{4k(k+\ell)}{3 \ell^2},~~~~\ell>2k
\eeq
for $(k,\ell)$ taken to be positive integers -- the inequality is required to keep $c\in(0,1)$ and violating this would change the topology of $\Sigma_4$. However this orbifold is fibred over S$^2$ so we need to take a bit more care. Notice that 
\beq
\frac{1}{2\pi}\int_{\text{S}^2}dD\psi= \ell,
\eeq
which is an integer, so this orbifold fibration is well defined at generic points in the space. There is however generically an issue at $x=x_0$, where the $\mathbb{R}^2/\mathbb{Z}_k$ orbifold fixed point lies. For this fixed point to be well defined the metric on its covering space should regular. The covering space is $\mathbb{R}^2$ fibred over S$^2$ where $\varphi= \frac{1}{k} \psi$ has period $2\pi$. We find that
\beq
\frac{1}{k}D\psi= d\varphi-\frac{\ell}{2k}\rho,
\eeq
so it follows that we should tune
\beq
\frac{\ell}{k}\in\mathbb{Z}.
\eeq
With the constants so tuned we see that $\Sigma_4$ has the topology of $\mathbb{CP}^2$ with two orbifold fixed points. If $\mathbb{CP}^2$ is parametrised as a squashed 3-sphere foliated over an interval, these are located at the boundaries of that interval. It is instructive to compute the Euler characteristic of the internal space. On any 4-manifold M$_4$ this is given through the formula
\begin{align}
\chi_E=\frac{1}{32\pi^2} \int_{\text{M}_4} (R_{abcd}R^{abcd}-4R_{ab}R^{ab}+R^2)\text{vol}(\text{M}_4). 
\end{align}
For the case at hand we find that
\beq
\chi_E(\Sigma_4)=\frac{(2k+\ell )}{k \ell}=3 - 2\left(1- \frac{1}{k}\right)  -\left(1- \frac{1}{\ell} \right).
\eeq 
This is consistent with the weighted projective space $\mathbb{WCP}^2_{[k,k,\ell]}$. A distinct version of $\mathbb{WCP}^2_{[k_1,k_2,k_3]}$, realised as an interval foliation of a U(1) orbifold fibration over $\mathbb{WCP}^1_{[k_1,k_2]}$,  was recently constructed in  \cite{Conti:2025rfd}, though this did not admit the tuning $k_1=k_2=k$. We find that
\beq
\frac{1}{2\pi}\int_{(x,\psi)}d{\cal A}= \frac{(\ell-2k)}{k \ell},
\eeq
which we note is consistent with one of the 4 possible supersymmetry preserving first Chern classes for the gauge field on $\mathbb{WCP}^2$ that generalise the twist and anti-twist on the spindle as discussed in \cite{Conti:2025rfd}.

We can uplift this solution to $d=11$ following the formulae in section \ref{eq:1011duplifts}, the total internal space is then S$^4/\mathbb{Z}_{\tilde{k}}$ fibred over $\Sigma^4$ in terms of ${\cal A}$, see \eqref{eq:11dmet}. As discussed around \eqref{eq:quantisationcondtion}, for this fibration to be well defined we need that the integrals of $\frac{\tilde{k} g}{2}d{\cal A}$ through the 2-cycles in $\Sigma_4$ are  quantised appropriately. In this case, the 2-cycles available to us are $(x,\psi)$ at constant values on the 2-sphere, and the 2-sphere itself at $x=x_0$. We find that the gauge field gives rise to charges
\beq
\frac{1}{2\pi}\int_{(x,\psi)}\frac{\tilde{k}g}{2}d{\cal A}= \frac{\tilde{k}(\ell-2k)}{2k \ell},~~~~~\frac{1}{2\pi}\int_{\text{S}^2,x=x_0}\frac{\tilde{k}g}{2}d{\cal A}=\frac{\tilde{k}(\ell-2k)}{2k}.
\eeq
A well defined fibration requires that the first of these should be quantised as an integer divided by $k \ell$ and the second as an integer. Given that we must have $\ell>2k$ and $\frac{\ell}{k}\in\mathbb{Z}$ for the $d=7$ solution to exist and be well defined we can fix
\beq
\ell=(2+q)k,~~~~q\in\mathbb{N}.\label{elltuning}
\eeq
In terms of this, both quantisation conditions are realised if
\beq
\tilde{k}q\in 2\mathbb{Z}.
\eeq
Which implies that either $\tilde{k}$ or $q$ must be an even number, and only  the later option allows the solution to be compatible with any value of $\tilde{k}$, including $\tilde{k}=1$ where there is no orbifold of the 4-sphere. Thus a well defined solution describing M5 brane wrapped on $\mathbb{WCP}^2$ exists but requires that the weights and possibly also $\tilde{k}$ are intricately tuned. We believe this awkward tuning of the integers and the need to take an orbifold of the 4-sphere in one instances is due to our solution being constructed in minimal $d=7$ gauged supergravity. Likely a generalisation exists in the U(1)$^2$ invariant sector of the maximal $d=7$ theory that is more permissive.  

We should also have that the $d=11$ flux $G$ is quantised over the two obvious 4-cycles it has support on: First off at constant values of the coordinates on $\Sigma_4$ we should have that
\begin{equation}
\frac{1}{(2 \pi)^3} \int_{\text{S}^4} G = N, 
\end{equation}
where $N$ is the number of M5 branes wrapping $\Sigma_4$. This fixes the constant $L$ in the lift to $d=11$ as $L^6 = \frac{ \tilde{k}^{2} N^{2} \pi^{2} }{8}$. 
Second, at the poles of the 4-sphere (where $\sin\beta =\pm 1$) we can integrate the 4-form over $\Sigma_4$ using \eqref{eq:4formoversigma4}, which yields
\beq
\mp \frac{1}{(2 \pi)^3} \int_{\Sigma_4^{\pm}} G= \frac{(\ell-2k)^2}{8 k^2 \ell}\tilde{k}N. 
\eeq
As this charge is defined  on an orbifold it is not entirely clear to us what the quantisation condition should be for this flux, however it is clearly possible to tune $N$ or $\tilde{k}$ such that it is integer. With the fluxes appropriately quantised we can now use \eqref{eq:holcent} to compute the central charge of the dual CFT$_2$ holographically, we find
\beq\label{eq:ccAdS3WCP2}
c_{\text{hol}}= \frac{(\ell-2k)^4}{8k^2\ell((\ell+2k)^2+2 \ell^2)}N^3 \tilde{k}^2,
\eeq
which is clearly rational, as one would expect on an orbifold. Likewise we can compute the R-symmetry Killing vector from \eqref{eq:Rsymmsugra}, given how $n$ is tuned in this section this yields
\beq \label{eq:RsymmWCP2sugra}
R=- \frac{8 \ell k(\ell+k)}{2\ell^2+(\ell+2k)^2}\partial_{\psi}.
\eeq
We will be able to match both of these results via a field theory computation in section \ref{eq:QFTWCP2}.

\subsection{On the preservation of global supersymmetry}
The way that we arrived at this solution guarantees that supersymmetry is preserved locally, but not that it is preserved globally. For this to be the case we need the Killing spinor on the 4 dimensional internal space to be well defined, which requires that (when ${\cal A}$ is also well defined) it is either periodic or anti-periodic under $\psi\to \psi+2\pi$ and regular on the covering spaces of the $\mathbb{R}^4/\mathbb{Z}_{\ell}$ and $\mathbb{R}^2/\mathbb{Z}_k$ orbifold fixed points. In terms of bi-linears, the periodicity condition is equivalent to the bi-linears in \eqref{eq:bispinors} being invariant under $\psi\to \psi+2\pi$, which in this case boils down to demanding this for the complex vielbein direction $W$ which contains the only dependence on $\psi$. The first thing to appreciate is that ${\cal A}$ is not currently well defined at the fixed points, but the supersymmetry conditions are invariant under
\beq
{\cal A}\to {\cal A}+ B d\psi,~~~~~W\to e^{-ig B\psi}W,\label{eq:transformations}
\eeq    
for $B$ a constant. We can use this fact to make ${\cal A}$ regular on the covering space of the orbifolds by fixing
\beq
4 g B=4b+\frac{3}{k}+\frac{4}{\ell}-\frac{3}{k+\ell}.
\eeq
This transformation leads to 
\beq
g{\cal A}\bigg\lvert_{x=x_0}=\frac{bk+1}{k}d\psi-\frac{\ell-2k}{2k}\rho,~~~~~g{\cal A}\bigg\lvert_{x=1}=\frac{b\ell +2}{\ell}d\psi,
\eeq
which gives a well defined gauge field when $b\in\mathbb{Z}$, as the $\psi$ dependence becomes pure gauge on the covering spaces of the two orbifold fixed points, where $\psi$ has periods $2\pi k$ and $2\pi\ell$ respectively. Before applying \eqref{eq:transformations} the $\psi$ dependence in $W$ is contained in the phase $e^{i n \psi}$, after applying it this phase is mapped to
\beq
e^{i n \psi}\to e^{- b \psi},
\eeq 
which does indeed make $W$ invariant under $\psi\to \psi+ 2\pi$. We can also most easily address regularity of the spinors on the covering spaces for the orbifold fixed points in terms of the bi-linears \eqref{eq:bispinors}: One can show that
\beq
x=x_0~~~\Rightarrow ~~~\cos\alpha=-1,~~~~~x=1~~~\Rightarrow~~~ \cos\alpha=1,
\eeq 
meaning that at these respective loci the bi-linears are simply spanned by the SU(2)-structure forms $(J_{\mp},\Omega_{\mp})$ as defined in \eqref{eq:su2forms}.

Close to $x=x_0$ when $k=1$ the space is perfectly regular so there is no potential issue with the spinor, however we do have an orbifold singularity at this point when $k> 1$ and $\ell=(2+q) k$. When this is the case, and after applying \eqref{eq:transformations}, we find that the canonical vielbein directions spanning $(J_-,\Omega_-)$ are
\beq
e^{-A}\bar{V}=\frac{3\ell (3\ell+2k)}{4(\ell+k)^{\frac{1}{2}}(\ell+2k)^{\frac{3}{2}}}\left(dz+\frac{i}{k}z D\psi\right),~~~~e^{-A}W=e^{-i b\psi}\frac{4k^2+4k\ell+3\ell^2}{2\sqrt{2}\sqrt{k \ell}(2k+\ell)}e^{\Delta}(dx_1+i dx_2).
\eeq
Close to $x=x_0$ we have that
\begin{align}
ds^2(\Sigma_4)&\to  Dy_i^2+\frac{(4k^2+4k\ell+3\ell^2)^2}{8k \ell(2k+\ell)^2}ds^2(\text{S}^2),~~~~g{\cal A}\to (bk+1)d\varphi-\frac{2k-\ell}{2 k} \rho\nn\\[2mm]
Dy_i&= dy_i+\frac{\ell}{2k} \epsilon_{ij}y_j\rho,~~~~y_1+i y_2= \frac{3\ell (3\ell+2k)}{4(\ell+k)^{\frac{1}{2}}(\ell+2k)^{\frac{3}{2}}}\sqrt{x}e^{i\varphi},~~~\varphi=\frac{1}{k}\psi.
\end{align}
The covering space at this point is then clearly $\mathbb{R}^2\hookrightarrow \hat{\Sigma}_4\to \text{S}^2$, where $\varphi$ has period $2\pi$, making the covering space regular when \eqref{elltuning} is imposed. The spinors the background supports are well defined at $x=x_0$ if on the covering space there exists a gauge in which they decompose in terms of constant spinors\footnote{More precisely we mean that this should be true in a frame in which the only dependence on $y_i$ is contained in $Dy_i$, which requires a rotation of the canonical complex vielbein component $V$. Further, this gauge should be one in which the spinor is not charged under the gauge field, \textit{i.e.} one in which ${\cal A}$ is $d\psi$ independent and is accessible via a globally well defined gauge transformation on the covering space.} on $\mathbb{R}^2$ and spinors on S$^2$.\footnote{See \cite{Ferrero:2021etw} for an in depth discussion on this point.} In terms of bi-linears this is equivalent to $(J_-,~\Omega_-)$ only depending on $y_i$ through $Dy_i$, which is automatic for $J_-$, but we find that 
\beq
\Omega_-=  \bar{V}\wedge W \propto e^{-i(1+k b) \varphi} (Dy_1+i Dy_2)\wedge e^{\Delta}(dz_1+i dz_2),
\eeq
clearly in addition to $Dy_i$ this also depends on $\varphi$, however so does ${\cal A}$ meaning that the spinors of the background are charge under the gauge field. We can turn off the $\varphi$ dependence by again making use of \eqref{eq:transformations}, this time for 
\beq
B=-\frac{1+b k}{g k}~~~~\Rightarrow~~~~{\cal A}\to {\cal A}-(1+bk)d\varphi,
\eeq
which on the covering space is a globally well defined gauge transformation. This also maps $\Omega_-\to  e^{i(1+k b) \varphi}\Omega_-$ which remove the $\varphi$ dependent phase of $\Omega_-$.  From this we see that the spinors on the covering space clearly do decompose in terms of constant spinors on $\mathbb{R}^2$ and as such are regular - it follows that supersymmetry is preserved at the $\mathbb{R}^2/\mathbb{Z}_k$ fixed point.

On the other hand, close to $x=1$ we have that the metric on $\Sigma_4$ and gauge field tend to
\begin{align}
ds^2(\Sigma_4)&\to \frac{(4k^2+3k\ell+3\ell^2)}{3\ell \sqrt{k(\ell+k)}}\left(dZ_1 d\overline{Z}_1+dZ_2 d\overline{Z}_2\right),~~~~ g{\cal A}= (b\ell +2)d\tilde{\varphi},~~~~\tilde{\varphi}=\frac{1}{\ell}\psi,\nn\\[2mm]
Z_1&=\sqrt{1-x}e^{i \tilde{\varphi}}\cos\left(\frac{\hat\theta}{2}\right),~~~~Z_2=\sqrt{1-x}e^{i (\tilde{\varphi}+\hat\phi)}\sin\left(\frac{\hat\theta}{2}\right),
\end{align}
where we have parametrised the 2-sphere as
\beq
e^{\Delta}=\frac{2}{1+z_1^2+z_2^2},~~~~z_1= \tan\left(\frac{\hat\theta}{2}\right)\cos\hat\phi,~~~~z_1= \tan\left(\frac{\hat\theta}{2}\right)\sin\hat\phi\label{eq:2sphereembedding},
\eeq
without loss of generality. Thus clearly the regular covering space for this orbifold is simply $\mathbb{R}^4$, where $\tilde{\varphi}$ has period $2\pi$. On the covering space the dependence of ${\cal A}$ on $\tilde{\varphi}$ is pure gauge, and can be turned off by operating with \eqref{eq:transformations} for
\beq
B=-\frac{b\ell +2}{g}.
\eeq
In this case the bi-linear are spanned by $(J_+,\Omega_+)$, which after applying the above gauge transformations tend to 
\begin{align}
J_+&\propto\frac{i}{2}(dZ_1\wedge d\overline{Z}_2+ d Z_2\wedge d\overline{Z}_2),~~~~\Omega_+\propto dZ_1\wedge dZ_2,
\end{align}
close to $x=1$ -- we have omitted only an overall constant. Clearly this behaviour is regular on the covering space, which can only happen if the spinors on M$_4$ are likewise regular there, so supersymmetry is also preserved at $x=1$.

\section{M5 branes wrapping spindles fibred over $\Sigma_{\mathfrak{g}}$}\label{sec:M5spindlesSigma}
In this section we construct solutions whose uplifts to $d=11$ describe M5 branes wrapping a spindle fibred over a genus $\mathfrak{g}$ Riemann surface. A similar solution, albeit strictly for $\mathfrak{g}>1$, was constructed in the U(1)$^2$ invariant sector of maximal $d=7$ gauged supergravity in \cite{Cheung:2022ilc} -- we recover a tuning of this in appendix \ref{eq:knownsols2}. In \cite{Cheung:2022ilc} a consistent truncation of the $d=7$ theory on a spindle down to minimal $d=5$ gauged supergravity was realised and then used to construct their fibred solution. None of the solutions we present can be constructed in this way, neither are we limited to $\mathfrak{g}>1$. Indeed we also find explicit solutions for which the base is both S$^2$ and $\mathbb{T}^2$.

A note of caution: The 3 solutions in each of the 3 following sections are all parametrised in terms of functions $(H,~P)$ - these are distinct functions for each solution following from different tunings of $(c_1,~c_2,~\kappa)$.\\
~\\
For all the three cases we will study in detail, we find that the central charge and the R-symmetry Killing vector can be written in a universal form. The R-symmetry Killing vector is given by taking \eqref{eq:Rsymmsugra} and substituting the proper value of $n$ according to the specific case we are interested in, we find
\begin{equation} \label{eq:RsymmsugraSigmakappa}
R = \frac{8 k_1 k_2 ( (k_1^2 + k_1 k_2 + k_2^2)q - 3 \kappa (k_1 + k_2)) }{8 \kappa (k_1^2 + k_1 k_2 + k_2^2) - 3 q (k_1 + k_2)(k_1^2 + k_2^2)} \partial_{\psi} ,
\end{equation}
For the holographic central charge we find that the three expressions we will derive by means of \eqref{eq:holcent}, can be written as
\begin{equation}\label{eq:hccSigmakappa}
c_{\text{hol}} = \frac{(k_2-k_1)^3 ( 24 \kappa^2 - 12 \kappa (k_1+k_2)q + (k_1^2+4 k_1 k_2+k_2^2)q^2)}{8 k_1 k_2 ( 3 (k_1+k_2)(k_1^2+k_2^2)q - 8 \kappa (k_1^2+k_1 k_2+k_2^2) ) } | |\kappa| \mathfrak{g} - 1 | \tilde{k}^2 N^3 ,
\end{equation}
where $k_1$ and $k_1$ are the two weights of the spindle with $k_2 > k_1$. For $\kappa=1$ we recover the result for S$^2$, for $\kappa=0$ yields that of $\mathbb{T}^2$ while for $\kappa=-1$ we recover that of $\Sigma_{\mathfrak{g}}$ with $\mathfrak{g}>1$. 
We now focus on the derivation and the analysis of these solutions.

\subsection{The 2-sphere}
We now turn our attention to constructing a solution with a spindle fibred over S$^2$. While it is possible to locally construct such a solution within the tuning of \eqref{eq:c2tunign}, this leads to the $D\psi$ fibre over S$^2$ being ill defined. We find it useful to instead fix
\beq
\kappa=1,~~~~c_1=\frac{c-1}{c},~~~~c_2= \frac{4b(c-1)^3(c+1)}{c},~~~~y =2(c-1)^2 x.
\eeq
We find then that the local solution becomes
\begin{align}
ds^2&=\frac{2}{g^2}\left(\frac{ x^{\frac{3}{2}}(1-c)^4H}{3(1-x)}\right)^{\frac{2}{5}}\bigg[ds^2(\text{AdS}_3)+(1-x)\left(ds^2(\tilde{\Sigma})+ \frac{3}{4(2-c)c x} ds^2(\text{S}^2)\right)\bigg],\nn\\[2mm]
ds^2(\tilde{\Sigma})&=\frac{9}{P}dx^2+\frac{n^2 P}{4xH^2}D\psi^2,~~~~D\psi=d\psi-\frac{1}{c n}\rho,~~~~d\rho=-\text{vol}(\text{S}^2),\nn\\[2mm]
g {\cal A}&=\frac{2(1-c^2)n(b+(x-2)x) n}{H}D\psi-(1-c) nd\psi,~~~~X^5= \frac{3(1-c)(1-x)x}{H},\label{eq:S2met}\\[2mm]
{\cal F}_4&=\frac{(1-c)^2(c+1)}{g^3}\left(\frac{2\sqrt{2}(b+(x-2)x)}{3 (x-1)^2} \text{vol}(\text{AdS}_3)\wedge dx+ \frac{n(b-x^2)}{2\sqrt{2}(c-2)cx^2}d\psi\wedge dx\wedge \text{vol}(\text{S}^2)\right),\nn
\end{align}
where $(H,P)$ are two functions of $x$ defined as
\beq
P=8(c+1)(2-c)x(b+(x-2)x)-H^2,~~~~H=(c+1)(b+(x-2)x)+2(c-2)x(x-1).
\eeq
The $x$ interval in the above solution is bounded as  $0<x_1<x<x_2<1$ for  $x_i$ the two smallest real roots of $P$ whenever the bounds
\beq
0<b<1,~~~~0<c<1,\label{eq:boundsS2}
\eeq
are obeyed. The roots themselves solve the equations
\beq
x_1(6-4(c-2)\sqrt{x_1}-3(c-1)x_1)=b(c+1),~~~~x_2(6+4(c-2)\sqrt{x_2}-3(c-1)x_2)=b(c+1),\label{eq:bx1x2}
\eeq
respectively, the consistency of which requires 
\beq
c= \frac{x_+^2(4+3 x_-)+ 3 x_-(2+x_-)^2}{x_+^2(2+3 x_-)+3 x_-^2(2+x_-)},~~~x_{\pm}=\sqrt{x_1}\pm \sqrt{x_2}, \label{eq:cx1x2}
\eeq
which in particular makes $x_-$ negative. The explicit analytic expressions for $(x_1,x_2)$ are rather complicated, we will not quote them explicitly as for now \eqref{eq:bx1x2} will be sufficient for our purposes - later we will also require \eqref{eq:cx1x2}. As $x\to x_i$ we find that
\beq
ds^2(\tilde{\Sigma})\propto  dz^2+  \frac{n^2 P'(x_i)^2}{144H(x_i)^2 x_i} z^2D\psi^2,
\eeq
for $x=x_i\pm z^2$ as appropriate. This gives an $\mathbb{R}^2/\mathbb{Z}_{k_i}$ orbifold singularity at $x=x_i$ provided that
\beq
\frac{n P'(x_1)}{12\sqrt{x_1}H(x_1)}=\frac{1}{k_1},~~~~\frac{n P'(x_2)}{12\sqrt{x_2}H(x_2)}=-\frac{1}{k_2},~~~~\psi\sim\psi+2\pi~~~~n>0\label{eq:needtobeintegersS2}
\eeq
which make $k_i$ positive. The orbifolds also require that $k_i$ are integer, which requires tuning $(b,c)$. We will simply assume such a tuning is possible for now, delaying a proof until we have collected more information.  Given the constraints \eqref{eq:bx1x2} we can eliminate $b$ in the above arriving at
\beq
k_1=\frac{1}{n(1+(1-c)\sqrt{x_1})},~~~~~k_2=\frac{1}{n(1-(1-c)\sqrt{x_2})}\label{eq:needtobeintegersS22},
\eeq
from which it follows that $k_2>k_1$. It is a simple matter to show that if we send $D\psi\to d\psi$ in $ds^2(\tilde{\Sigma})$, then
\beq
R(\tilde{\Sigma})\text{vol}(\tilde{\Sigma})= \frac{n}{6}\partial_{x}\left(\frac{2x PH'+H'(P-xP')}{x^{\frac{3}{2}}H^2}\right),
\eeq
from which it follows that the Euler characteristic of $\tilde{\Sigma}$ is
\beq
\chi_E= \frac{1}{4\pi}\int_{\tilde{\Sigma}}R(\tilde{\Sigma})\text{vol}(\tilde{\Sigma})=(2+(c-1)(\sqrt{x_2}-\sqrt{x_1}))n=\frac{k_1+ k_2}{k_1k_2},
\eeq
where we have used \eqref{eq:bx1x2} in the first equality to eliminate $b$ at each bound of $x$, then \eqref{eq:needtobeintegersS22} in the second equality. Likewise we find that
\beq
\frac{g}{2\pi}\int_{\tilde{\Sigma}}d{\cal A}=(c-1)n(\sqrt{x_1}+\sqrt{x_2})=\frac{(k_1-k_2)}{k_1 k_2}\label{eqfirstspin},
\eeq
meaning that $\tilde{\Sigma}=\mathbb{WCP}^1_{[k_1,k_2]}$, a spindle, and the field strength of ${\cal A}$ is appropriately quantised on this space in terms of an anti-twist. In the full space $\tilde{\Sigma}$ is fibred over S$^2$ so we should also have that
\beq
\frac{1}{2\pi}\int_{\text{S}^2}dD\psi=\frac{2}{c n}\in \mathbb{N}.
\eeq
Likewise, upon lifting to $d=11$ as in section \ref{eq:1011duplifts} we have to ensure that
\beq
\frac{g \tilde{k}}{2}\frac{1}{2\pi}\int_{\text{S}^2,x=x_i}d{\cal A}= \frac{\tilde{k}}{2}\left(\frac{2}{c n k_i}-2\right) \in\mathbb{Z},
\eeq
and that $\frac{\tilde{k}}{2}\times$\eqref{eqfirstspin} is an integer divided by $k_1 k_2$, so that the fibration of S$^4/\mathbb{Z}_{\tilde{k}}$ over the metric in \eqref{eq:S2met} is well defined. To this end we find it helpful to tune 
\beq
c=\frac{2}{n k_1 k_2 q},~~~~q> \frac{2}{k_1},~~~~~q\in\mathbb{N}, \label{eq:cineq}
\eeq
with the inequality required to keep $c$ in the range of $(0,1)$ given \eqref{eq:needtobeintegersS22}, this leads to
\beq
\frac{1}{2\pi}\int_{\text{S}^2}dD\psi = q k_1 k_2,~~~~\frac{g}{2\pi}\int_{\text{S}^2,x=x_1}d{\cal A} =q k_2-2,~~~~\frac{g}{2\pi}\int_{\text{S}^2,x=x_2}d{\cal A} =q k_1-2,
\eeq
which are all positive due to the inequality in \eqref{eq:cineq}. Thus the fibration of S$^4/\mathbb{Z}_{\tilde{k}}$ over $\tilde{\Sigma} \times\text{S}^2$  is well defined if
\beq
(q,~k_2-k_1)\in 2\mathbb{N},~~~~\text{or}~~~~  \tilde{k}\in 2\mathbb{Z},
\eeq
implying that either $q$ is even and both $(k_1,k_2)$ are odd or $\tilde{k}$ is even - only the former  choice is compatible with an embedding in terms of the round 4-sphere.

We are now in a position to prove that we can indeed take $(k_1,k_2)$ to be integers obeying $k_2>k_1>0$. To achieve this we require expressions for $x_{\pm}=\sqrt{x_1}\pm \sqrt{x_2}$ in terms of $(k_1,k_2,q)$. Fortuitously  \eqref{eq:cx1x2}, \eqref{eq:needtobeintegersS22} and \eqref{eq:cineq} provides us with what we need, we find
\begin{align}\label{eq:xpmexp}
x_+ & = \frac{4q(3 (k_1^2- k_2^2) + q(k_2^3-k_1^3))}{24(k_1+k_2)-16(k_1^2+k_1 k_2+k_2^2)q+3q^2(k_1+k_2)(k_1^2+k_2^2)}, \nn \\[2mm]
x_- & = - \frac{2q (k_1-k_2)^2(q(k_1+k_2)-2)}{24(k_1+k_2)-16(k_1^2+k_1 k_2+k_2^2)q+3q^2(k_1+k_2)(k_1^2+k_2^2)},
\end{align}
which means we can now also express $(b,~c,~n)$ in terms of $(k_1,~k_2,~q)$ by using \eqref{eq:bx1x2}, \eqref{eq:cx1x2} and \eqref{eq:cineq}. These expressions are long winded, so we will not quote them explicitly but we have checked that $c,b\in(0,1)$, as the existence of this solution demands, provided that
\beq
k_2>k_1>0,~~~~ q>\frac{2}{k_1}.
\eeq
There is no further requirement on these constants, and so no issue with taking them all to be integers that obey these bounds.

With \eqref{eq:xpmexp} in hand we can now compute the charges of the $d=11$ flux. As before, we can tune $L$ such that the charge on $\text{S}^4$ is $N$. Using the formula \eqref{eq:4formoversigma4} we then find that the charge of the 4-form on $\tilde{\Sigma}\times \text{S}^2$ at $\sin\beta=\pm 1$ is given by 
\beq
\mp \frac{1}{(2\pi)^2}\int_{\tilde{\Sigma}\times \text{S}^2}G=\frac{(k_2-k_1)(q(k_1+k_2)-4)}{8 k_1 k_2}\tilde{k}N,
\eeq 
which is again rational in general, but can be made integer by tuning $\tilde{k} N$. On the other hand, we can compute the holographic central charge of this solution using \eqref{eq:holcent}, we find 
\beq\label{eq:hccspindlesS2}
c_{\text{hol}}=\frac{(k_2-k_1)^3(q^2(k_1^2+4k_1 k_2+k_2^2)-12q(k_1+k_2)+24)}{8 k_1 k_2(3q(k_1+k_2)(k_1^2+k_2^2)-8 (k_1^2+k_1 k_2+k_2^2))}\tilde{k}^2 N^3,
\eeq
yielding another rational result. 

\subsubsection{On the preservation of global supersymmetry}\label{eq:sphereglobalsym}
We now turn our attention to the preservation of global supersymmetry: Having established that it is indeed possible to take $(k_1,k_2)$ to be positive integers constrained as $k_2>k_1$, we would now like to additionally impose  $\text{gcd}(k_1,k_2)=1$, then there exist integers $(m_1,m_2)$ such that
\beq
k_1m_2-k_2m_1=1,
\eeq
due to B\'ezout's identity. Under this assumption we can perform a local gauge transformation such that the gauge field and phase $e^{i n \psi}$ in the vielbein direction $W$ are mapped to
\beq
{\cal A}\to {\cal A}+ \left(\frac{m_1(k_1-k_2)+n k_1-1}{g k_1}\right) d\psi,~~~~ e^{ i n\psi}\to e^{i(m_2-m_1)\psi},\label{eq:S2gaguetrans1}
\eeq 
which make the spinors the solution supports well behaved under $\psi\to \psi +2\pi$ and is such that as $x\to x_i$ at constant values on S$^2$
\beq
g{\cal A}\to \frac{m_i(k_1-k_2)}{k_i}d\psi,
\eeq
making the gauge field regular on the covering space of the  $\mathbb{R}^2/\mathbb{Z}_{k_i}$ fixed points. Thus supersymmetry is preserved globally if the spinors the background supports are regular at the poles of the spindle - as before this is most easy to address in terms of the bi-linears in \eqref{eq:bispinors}: This time we find that
\beq
x=x_1~~~\Rightarrow ~~~\cos\alpha=-1,~~~~~x=x_2~~~\Rightarrow~~~ \cos\alpha=1,
\eeq 
so again \eqref{eq:bispinors} are simply spanned by the SU(2)-structure forms $(J_{\mp},\Omega_{\mp})$ at the respective poles. Close to $x=x_j$ we find that
\begin{align}
ds^2(\tilde{\Sigma})\propto  (Dy^{(j)}_a)^2,~~~g{\cal A}= m_j (k_2-k_1) d\psi_j+...,~~~~\psi_j=  \frac{\psi}{k_j}\nn\\[2mm]
Dy^{(j)}_a= dy^{(j)}_a+ \frac{k_1 k_2}{2 k_j}\epsilon_{ab}y^{(j)}_b \rho,~~~y^{(j)}_1+iy^{(j)}_2 =r e^{i\psi_j}
\end{align}
where there is no summation over $j$, which is regular on the covering spaces of the $\mathbb{R}^2/\mathbb{Z}_{k_j}$ fixed points where $\psi_j$ has period  $2\pi$. Notice that on the covering spaces  the $d\psi_j$ dependence in ${\cal A}$ is pure gauge, so can be turned off entirely by operating with \eqref{eq:transformations}. The spinor is well defined at the poles of the spindle if it is regular on the covering space in such a gauge. This in turn is equivalent to $(J_{\mp},\Omega_{\mp})$ only depending on $(r,\psi)$ through $(y^{(j)}_1,y^{(j)}_2)$ at the respective poles, which in-fact follows if the same is true of (where $V$ is the canonical vielbein direction of section \ref{sec:class2}) 
\beq
{\cal V}=\left\{\begin{array}{l}e^{i n\psi} V,~~~~x=x_2\\[2mm]
                           e^{i n\psi} \overline{V},~~~~x=x_1 \end{array}\right. ,
\eeq 
after first applying \eqref{eq:S2gaguetrans1}, expanding about the poles and then gauging away the $\psi_j$ dependence. We find that this chain of operations leads to
\beq
{\cal V}\to {\cal C}\bigg(  d(y^{(j)}_1+i y^{(j)}_2)- i \frac{k_1 k_2 q}{2 k_j} (y^{(j)}_1+i y^{(j)}_2)\rho\bigg)
\eeq 
at $x=x_j$,  where again there is no summation over $j$, and ${\cal C}$ is a non-zero constant. This implies that the spinors the background supports are regular on the covering spaces of the orbifolds and so it follows that supersymmetry is globally preserved.

\subsection{$\Sigma_{\mathfrak{g}}$ with $\mathfrak{g}>1$ but otherwise free}
In this section we construct a solution with a spindle fibred over a genus $\mathfrak{g}>1$ Riemann surface, with $\mathfrak{g}$ otherwise free. A simpler solution of this type also exists within the tuning of \eqref{eq:c2tunign} which we present in  appendix \ref{sec:simple H_2} but, similar to the solution in \cite{Cheung:2022ilc},  this requires one to tune $\mathfrak{g}$ to have a well defined solution.

To extract this global completion we find it helpful to tune the coefficients in \eqref{eq:CCsol3}, $\kappa$ and the $y$ coordinate as
\beq
c_1= \frac{1}{c-1},~~~~c_2=-\frac{4b(c-2)}{c-1},~~~~\kappa=-1,~~~~y= \frac{2x}{c^2},
\eeq
after which the local solution can be written as
\begin{align}
ds^2&=\frac{2}{g^2}\left(\frac{x^{\frac{3}{2}}H}{3(1-x)}\right)^{\frac{2}{5}}\bigg[\frac{1}{c^2}ds^2(\text{AdS}_3)+(1-x)\left(ds^2(\tilde{\Sigma})+ \frac{3}{4(1-c^2)x}ds^2(\Sigma_\mathfrak{g})\right)\bigg],\nn\\[2mm]
ds^2(\tilde{\Sigma})&=\frac{9}{P}dx^2+ \frac{n^2 P}{4 c^2 x H^2}D\psi^2,~~~~D\psi=d\psi+\frac{c}{(c-1)n}\rho,~~~~d\rho=-\text{vol}(\Sigma_{\mathfrak{g}}),\nn\\[2mm]
g{\cal A}&= \frac{2n F }{c H}D\psi-\frac{n}{c}d\psi,~~~X^5=\frac{3(1-x)x}{H},\label{eq:H2gensol}\\[2mm]
{\cal F}_4&=\frac{1}{c g^3}\left(\frac{2 \sqrt{2}F}{3 c^2(x-1)^2} \text{vol}(\text{AdS}_3)\wedge dx+ \frac{n(c^3(c-2)b+(2c-1)x^2)}{2 \sqrt{2}(1 - c^2)x^2}dx\wedge d\psi\wedge \text{vol}(\Sigma_{\mathfrak{g}})\right),\nn
\end{align}
where we define the functions
\begin{align}
F & =c^3(2-c)b + x(2c-1)(x-2),~~~~H=2(c+1)x(1-x) + F, \nn \\[2mm]
P & = 8 (c+1)x F-H^2
\end{align}
and have taken the quotient  $\mathbb{H}^2\to \mathbb{H}^2/\Gamma=\Sigma_{\mathfrak{g}}$ such that $\Sigma_{\mathfrak{g}}$ is a genus $\mathfrak{g}$ Riemann surface with $\mathfrak{g}>1$ and we take
\beq
\frac{1}{2\pi}\int_{\Sigma_{\mathfrak{g}}}\text{vol}(\Sigma_{\mathfrak{g}})=2( \mathfrak{g}-1),~~~~d\rho=-\text{vol}(\Sigma_{\mathfrak{g}}).\label{eq:H2quotient}
\eeq
The function $P$ is an order 4 polynomial which has at least 2 real (order 1) roots when the constants $(c,b)$ are constrained as
\beq
0<c< 1,~~~~0<b<1.\label{eq:H2consbc}
\eeq
Taking $(x_1,x_2)$ to be the smallest of these two real roots and assuming $x_2>x_1$ the above bound also ensures that $0<x_1<x_2<0$, as such it is simple to see that the $x$ interval is bounded as $x\in [x_1,~x_2]$ where only $P$ is vanishing at the bounds- the remaining functions defining the solution are just constant at these loci. Being that they come from an order 4 polynomial clearly the explicit expressions for the roots in terms of  $(c,b)$ are a mess, they do however obey
\beq
2(2-c)c^3=6c x_1+4(c+1)x_1^{\frac{3}{2}}+3 x_1^2,~~~~2(2-c)c^3=6c x_2-4(c+1)x_2^{\frac{3}{2}}+3 x_2^2.\label{eq:H2casebc}
\eeq
Having learnt a lesson during the analysis of the previous section, rather than working with $(x_1,~x_2)$ directly, we shall work in terms of
\beq
x_{\pm}= \sqrt{x_1}\pm \sqrt{x_2},
\eeq
which contain the same information, we can then use \eqref{eq:H2casebc} to fix $(c,b)$ as
\begin{align}
c&=-\frac{6x_-^2+3 x_-^3+2 x_+^2+3 x_- x_+^2}{2(6 x_-+3 x_-^2+x_+^2)},\nn\\[2mm]
b&=\frac{(x_+^2-x_-^2)^2(8 +6 x_-+x_-^2-x_+^2)(6x_-+3 x_-^2+x_+^2)^3}{(x_-+2)(4x_-+x_-^2+x_+^2)(6x_-^2+3x_-^3+2x_+^2+3 x_- x_+^2)^3}.\label{eq:cbH2}
\end{align}

Clearly as things stand the internal space of \eqref{eq:H2gensol} is behaving locally like a 2-sphere fibred over $\Sigma_{\mathfrak{g}}$, as before the global behaviour depends on what exactly is happening at $x=x_i$ for $i=1,2$.  One can easily show that $\tilde{\Sigma}$ has $\mathbb{R}^2/\mathbb{Z}_{k_i}$ orbifold fixed points at $x=x_i$ provided that $x_{\pm}$ obey
\beq
\frac{k_1+k_2}{k_1k_2}=\frac{n(2c+x_-)}{c},~~~~\frac{k_2-k_1}{k_1 k_2}= \frac{n x_+}{c}\label{eq:k1k2H2}
\eeq
and $(k_1,k_2)$ are taken to be integers. This also leads, at constant values on $\Sigma_{\mathfrak{g}}$, to
\beq
\chi_E(\tilde{\Sigma})=\frac{k_1+k_2}{k_1 k_2},~~~~-\frac{g}{2\pi}\int_{\tilde{\Sigma}}d{\cal A}=\frac{k_2-k_1}{k_1 k_2},
\eeq
so the internal space is in fact a spindle fibred over $\Sigma_{\mathfrak{g}}$ and the gauge field is appropriately defined such that we have an anti- twist. Given this topology we need that the connection that mediates the fibration is well defined which requires
\beq
\frac{1}{2\pi}\int_{\Sigma_{\mathfrak{g}}}dD\psi \in\mathbb{Z},
\eeq
which in principle already gives us enough constraints to fix $(c,b,n,x_{\pm})$ in terms of physical quantities. However, upon lifting to $d=11$ as in section \ref{eq:1011duplifts}, we also need the fibration of S$^4/\mathbb{Z}_{\tilde{k}}$ over $\tilde{\Sigma}\times\Sigma_{\mathfrak{g}}$ to be well defined - this requires the integrals of $\frac{g \tilde{k}}{2}d{\cal A}$ to be appropriately quantised over both $\tilde{\Sigma}$ and $\Sigma_{\mathfrak{g}}$ at $x=x_i$. To ensure this we first tune\footnote{Below we could of course fix $q=\frac{\tilde{q}}{\mathfrak{g}-1}$ and demand that $\tilde{q}$ is integer, but this complicates later expression.}
\beq
n=\frac{2c}{(1-c)k_1 k_2 q},~~~~ q (\mathfrak{g}-1)\in\mathbb{Z},\label{eq:ncondH2}
\eeq  
upon which we find 
\beq
\frac{1}{2\pi}\int_{\Sigma_{\mathfrak{g}}}dD\psi=q k_1 k_2 (\mathfrak{g}-1),~~~~\frac{g}{2\pi}\int_{\Sigma_{\mathfrak{g}}|~x=x_i}d{\cal A}=\left(2+\frac{k_1 k_2 q}{k_i}\right)(\mathfrak{g}-1), 
\eeq
which are all integer valued provided that $(k_1,k_2,q (\mathfrak{g}-1))$ are like-wise all integers. The quantisation condition such that the embedding into S$^4/\mathbb{Z}_{\tilde{k}}$ is well defined then amounts to imposing one of 
\beq
(k_2-k_1,q (\mathfrak{g}-1))\in 2\mathbb{Z}~~~~\text{or}~~~~ \tilde{k}\in 2 \mathbb{Z},
\eeq
similar to the case of the sphere only now the first option, which is compatible with a round 4-sphere embedding, can be achieved  with $(k_1,k_2)$ both odd and either $q$ even or $\mathfrak{g}>1$ odd.

We should now establish whether taking $(k_1,k_2,q)$ to be integers is compatible with the bounds on $(c,b)$ that are necessary for a solution to exist. We can establish this by simply solving \eqref{eq:cbH2}, \eqref{eq:k1k2H2} and \eqref{eq:ncondH2} to give us $x_{\pm}$ in terms of $(k_1,k_2,q)$, we find that
\begin{align}
x_+&= \frac{4q (3 ( k_2^2 - k_1^2) + q (k_2^3-k_1^3))}{24 (k_1+k_2)+16 (k_1^2+ k_1 k_2+k_2^2)q+3 q^2(k_1+k_2)(k_1^2+k_2^2)},\nn\\[2mm]
x_-&= - \frac{2(k_1-k_2)^2q(2+q(k_1+k_2))}{24 (k_1+k_2)+16 (k_1^2+ k_1 k_2+k_2^2)q+3 q^2(k_1+k_2)(k_1^2+k_2^2)},
\end{align}
then through \eqref{eq:cbH2} and \eqref{eq:ncondH2} we can extract expressions for $(c,b,n)$ in terms of $(k_1,k_2,q)$ also, that we will not quote explicitly. In terms of conditions on the weights and $q$ the constraints of \eqref{eq:H2consbc} and $0<x_1<x_2<1$ are then equivalent to
\beq
k_2>k_1>0,~~~~q>0,
\eeq
so there is no obstruction to taking $(k_1,k_2)$ integer provided the above bounds are satisfied.

When this solution is lifted to $d=11$ we should also have that the 4-form $G$ gives rise to quantised charges over S$^4$ and $\tilde{\Sigma}\times\Sigma_{\mathfrak{g}}$ at the poles of the 4-sphere where $\sin\beta=\pm 1$. As before we can simply tune $L$ such that flux on S$^4$ is consistent with $N$ M5 branes wrapping $\tilde{\Sigma}\times\Sigma_{\mathfrak{g}}$. The other charge can be computed through \eqref{eq:4formoversigma4} and gives rise to 
\beq
\mp\frac{1}{(2\pi)^3} \int_{\tilde{\Sigma}\times\Sigma_{\mathfrak{g}}}G= \frac{(k_2-k_1)(4+q(k_1+k_2))}{8 k_1 k_2}\tilde{k}N  (\mathfrak{g}-1),
\eeq
which again gives us a generically rational charge that can be made integer if we choose by tuning $\tilde{k}N $. The holographic central charge can then be computed through \eqref{eq:holcent} yielding
\beq\label{eq:hccSpindlesSigmag}
c_{\text{hol}}= \frac{(k_2-k_1)^3(24+12(k_1+k_2)q+q^2(k_1^2+4 k_1 k_2+k_2^2))}{8 k_1 k_2(8(k_1^2+k_1 k_2+k_2^2)+3 q (k_1+k_2)(k_1^2+k_2^2))}\tilde{k}^2 N^3(\mathfrak{g}-1),
\eeq 
giving a rational result as one would expect for a CFT compactified on an orbifold.

As in the previous solution, providing we assume that $\text{gcd}(k_1,k_2)=1$ as we are free to do, then there exist integers $(m_1,m_2)$ obeying $k_1m_2-k_2m_1=1$. Under this assumption it is possible to perform a gauge transformation such that the $d\psi$ leg of the gauge field and phase in $W$ are mapped precisely as in the previous section, in this case we must operate as in \eqref{eq:transformations} for
\beq
B= \frac{m_1(k_2-k_1)+ n k_1-1}{g k_1},
\eeq 
after which the gauge field is regular on the covering spaces of the $\mathbb{R}^2/\mathbb{Z}_{k_i}$ fixed points and the spinor the background supports transforms appropriately under $\psi\to \psi+2\pi$. One can also show that the spinor is well defined at the poles of the spindle in much the same fashion as in \ref{eq:sphereglobalsym}.
 
\subsection{The 2-torus}
Our final $d=4$ orbifold is a spindle fibred over $\mathbb{T}^2$, for which it is convenient to tune 
\beq
c_1=-\frac{b}{4},~~~~c_2= - b c,~~~~ y=2x,~~~~\kappa=0.
\eeq
The local form of the solution may then be expressed as
\begin{align}
ds^2&=\frac{2\left(\frac{x^{\frac{3}{2}}H}{3(1-x)}\right)^{\frac{2}{5}}}{g^2}\bigg[ds^2(\text{AdS}_3)+ (1-x)\bigg(ds^2(\tilde{\Sigma})+ \frac{3 b}{32 x}ds^2(\mathbb{T}^2)\bigg)\bigg],~~~X^5=\frac{3(1-x)x}{H},\nn\\[2mm]
ds^2(\tilde{\Sigma})&=\frac{9}{P}dx^2+\frac{n^2 P}{4x H^2},~~~D\psi=d\psi-\frac{b}{4n}\rho,~~~g{\cal A}=\frac{2n(c+(x-2)x)}{H}D\psi- \frac{n(c+2x-3x^2)}{H}d\psi,\nn\\[2mm]
g^3{\cal F}_4&=\frac{2\sqrt{2}(c+(x-2)x)}{3(x-1)^2}\text{vol}(\text{AdS}_3)\wedge dx+\frac{n b(x^2-c)}{16 \sqrt{2}x^2}dx\wedge d\psi\wedge \text{vol}(\mathbb{T}^2),
\end{align}
where this time the functions are given by
\beq
P=16x(c+(x-2)x)-H^2,~~~~H=c+(2-3x)x,
\eeq
such that $b$ just appears as a constant multiplying $\mathbb{T}^2$. The function $P$ is again an order 4 polynomial which has two complex roots and two real roots $(x_1,x_2)$ such that $0<x_1<x_2<1$ provided that $0<c<1$. Again the explicit expressions for the roots are a bit unwieldy and it is more convenient to work with $x_{\pm}= \sqrt{x_1} \pm \sqrt{x_2}$, which must obey 
\beq
x_+^2=-\frac{3 x_-(x_-+2)^2}{4+3 x_-},~~~~4c=-\frac{x_-(8+2x_-)(6+8 x_-+2 x_-^2)^2}{(4+3 x_-)^2},\label{eq:xpmctor}
\eeq 
when $(x_1,x_2)$ are roots.

The analysis of the topology of the space is quite similar to the previous section, indeed at the points $x=x_i$ for $i=1,2$ we find that the internal space tends to $\mathbb{R}^2/\mathbb{Z}_{k_i}$ fibred over $\mathbb{T}^2$ and\footnote{By $\chi_E(\tilde{\Sigma})$ we mean the Euler characteristic computed at constant values on $\mathbb{T}^2$}, 
\beq
\chi_E(\tilde{\Sigma})= \frac{k_1+k_2}{k_1 k_2},~~~~-\frac{1}{2\pi}\int_{\tilde{\Sigma}}d{\cal A}=\frac{k_2-k_1}{k_1 k_2},
\eeq
provided that we tune
\beq
(2+x_-)n=\frac{k_1+k_2}{k_1 k_2},~~~x_+ n=  \frac{k_1-k_2}{k_1k_2} \label{eq:xpmfix}.
\eeq
These indicate that the full internal space is a spindle fibred over $\mathbb{T}^2$ provided that we are free to take $(k_1,k_2)$ to be integers.
The conditions \eqref{eq:xpmctor} and \eqref{eq:xpmfix} imply that
\begin{align}
x_+&= \frac{4(k_2^3-k_1^3)}{3(k_1+k_2)(k_1^2+k_2^2)},~~~x_-=-\frac{2(k_2-k_1)^2}{3(k_1^2+k_2^2)},~~~n=\frac{3(k_1+k_2)(k_1^2+k_2^2)}{4k_1 k_2(k_1^2+k_1 k_2+k_2^2)}\nn\\[2mm]
c&=\frac{(k_2-k_1)^2(3k_1^2+2k_1 k_2+k_2^2)^2(k_1^2+2k_1 k_2+3k_2^2)^2(3k_1^2+2k_1k_2+3k_2^2)}{27(k_1+k_2)^4(k_1^2+k_2^2)^4},
\end{align}
which are indeed consistent with $0<c<1$ and $x_1<x_2$ provided that $k_2>k_1>0$ - there is no further constraint required to achieve this so we are free to take $(k_1,k_2)$ to be (positive) integers as the spindle interpretation demands.

We also need that the orbifold fibrations of both the spindle over $\mathbb{T}^2$, and upon lifting to $d=11$ as in  section \ref{eq:1011duplifts}, S$^4/\mathbb{Z}_{\tilde{k}}$ over $\tilde{\Sigma}\times \mathbb{T}^2$ are well defined, this leads us to tune
\beq
b= \frac{8n k_1 k_2 \pi q}{\text{Vol}(\mathbb{T}^2)},
\eeq
such that 
\beq
\frac{1}{2\pi}\int_{\mathbb{T}^2}dD\psi= 4 k_1 k_2 q,~~~~\frac{1}{2\pi}\int_{\mathbb{T}^2|x=x_i}d{\cal A}=\frac{k_1 k_2 q}{k_i},
\eeq
which are all integers provided that $q$ is, under our previous assumptions. Once more we must further demand that either  
\beq
(k_2-k_1,q)\in 2\mathbb{Z}~~~~\text{or}~~~~\tilde{k}\in 2\mathbb{Z},
\eeq
to have a well defined fibration in $d=11$ with only the former choice being compatible with a round 4-sphere embedding.

In $d=11$  we should also have that the flux of the 4-form $G$ is quantised over the 4-cycles available to us, as before we can tune the constant $L$ such that we have $N$ units of flux on the 4-sphere. The other cycle is $\tilde{\Sigma}\times \mathbb{T}^2$ at the poles of the 4-sphere where $\sin\beta=\pm 1$, this leads to
\beq
\mp \frac{1}{(2\pi)^3}\int_{\tilde{\Sigma}\times \mathbb{T}^2}G=\frac{k_2^2-k_1^2}{8 k_1 k_2} \tilde{k}N q,
\eeq
which is again rational in general, but can be made integer by tuning $\tilde{k}N$. The holographic central charge on the other hand takes the form
\beq\label{eq:hccSpindleT2}
c_{\text{hol}}= \frac{(k_2-k_1)^3(k_1^2+4k_1k_2+k_2^2)}{24 k_1 k_2(k_1+k_2)(k_1^2+k_2^2)} q\tilde{k}^2 N^3.
\eeq

Finally we address the issue of global supersymmetry preservation: A well defined geometry requires us to assume that $k_2>k_1$ are positive integers, if we further assume that $\text{gcd}(k_1,k_2)=1$, then we know that integers $(m_1,m_2)$ exists such that $k_1 m_2-k_2 m_1=1$. Under this assumption we can perform a local gauge transformations such that
\beq
g{\cal A}\to g{\cal A}+\left(\frac{m_1x_+}{1-\sqrt{x_2}}-n \sqrt{x_1}\right)d\psi,~~~~~ e^{i n\psi}\to  e^{i(m_2-m_1)\psi}. 
\eeq
As before this ensures that the spinors the background supports have a well defined periodicity under $\psi\to \psi+2\pi$ and ensures that at constant values on $\mathbb{T}^2$
\beq
g{\cal A}(x=x_i)=-\frac{m_i(k_2-k_1)}{k_i}d\psi,
\eeq
making the gauge field regular on the covering space of the $\mathbb{R}^2/\mathbb{Z}_{k_i}$ orbifold fixed points, and as such well defined. Finally, with a practically identical computation to section \ref{eq:sphereglobalsym}, one can establish that the spinor are regular at the poles of the spindle - as such supersymmetry is preserved globally.

\section{Field theory compactified on M$_4$}  \label{eq:fieldtheory}
In this section we wish to recover the holographic central charge of the geometries of the preceding 2 sections from a field theory computation. Fortuitously there is a well defined prescription for doing this which starts with the anomaly polynomial of the  ${\cal N}=(2,0)$ theory associated to $N$ M5 branes \cite{Witten:1996hc,Alday:2009qq,Hosseini:2020vgl}, and employs c-extremisation \cite{Benini:2012cz}\footnote{In the process of working out how to perform this computation we found \cite{Ferrero:2021wvk,Ferrero:2021etw,Ferrero:2020laf,Cheung:2022ilc,Arav:2022lzo,Bah:2011vv,Bah:2012dg,Benini:2012cz,Benini:2013cda,Benini:2015bwz,Hosseini:2021fge,Boido:2021szx,Boido:2022mbe,Amariti:2023gcx,Amariti:2023mpg} very useful.}. Allow us to describe this procedure in general terms before applying it in the following sub-sections.

We wish to compactify the $d=6$ theory of the M5 brane on both M$_4=\mathbb{WCP}^2_{[k,k,\ell]}$ and $\mathbb{WCP}^1_{[k_1,k_2]}\hookrightarrow \text{M}_4\to \Sigma_{\mathfrak{g}}$. Given that our geometric solutions are in minimal $d=7$ supergravity, which is itself a sub-sector of the U(1)$^2$ invariant sector of $d=7$ maximal, we should perform this compactification in the presence of two non-trival U(1) background gauge fields $(\hat A_1,\hat A_2)$ embedded in the Cartan subgroup of Sp(2), the R-symmetry of the ${\cal N}=(2,0)$ theory in $d=6$. The normal bundle to M5 branes wrapped on $\mathbb{R}^{1,1}\times \text{M}_4$ is then 
\beq
{\cal N}=\mathbb{R}\oplus {\cal N}_1\oplus {\cal N}_2,
\eeq    
for ${\cal N}_{1,2}$ complex line bundles. Given this, following \cite{Hosseini:2020vgl}, the 8-form anomaly polynomial for $N$ M5 branes compactified on M$_4$ takes, to leading order in $N$, the form 
\beq
{\cal A}_{6d}= \frac{1}{24} c_1({\cal N}_1)^2c_1({\cal N}_2)^2 N^3,
\eeq
where $c_1({\cal N}_{1,2})$ are the first Chern classes of the respective bundles. The effect of backreacting M5 branes on ALE spaces leads to a ${\cal N}=(1,0)$ theory in $d=6$ and was considered in \cite{Ohmori:2014kda}. For the case of $\text{S}^4/\mathbb{Z}_{\tilde{k}}$, which is relevant here, this amounts to modifying the anomaly polynomial as 
\beq
{\cal A}_{6d}= \frac{1}{24}  c_1({\cal N}_1)^2c_1({\cal N}_2)^2N^3\tilde{k}^2,\label{eq:6danomolypolynomial}
\eeq
which of course simply recovers the previous result when $\tilde{k}=1$ as it should. We wish to compactify the $d=6$ anomaly polynomial on an 8 dimensional space $Z_8$ of the form 
\beq
\text{M}_4 \hookrightarrow  Z_8 \to Z_4,
\eeq 
where $Z_4$ is the 4 manifold on which the anomaly polynomial of the ${\cal N}=(2,0)$ SCFTs in $d=2$ dual to our geometries will be defined on, after compactification. Given how our solutions are embedded into the U(1)$^2$ sector of maximal $d=7$ gauged supergravity, see section \ref{eq:1011duplifts}, and following  \cite{Cheung:2022ilc}, we  should be first identifying the background gauge fields of the field theory with the gauge field of our gravitational solutions as
\beq
\hat{A}_1=\hat{A}_2= \frac{g}{2} {\cal A}.
\eeq
When compactifying one needs to take account of the isometries of the compactification manifold M$_4$ - here the details depend on the specific M$_4$ we are talking about, but in all cases there is at least a U(1)$^2$ isometry spanned by two coordinates $(\partial_{\hat{\phi}},\partial_{\psi})$. We will refer to this symmetry as U(1)$_{J_1}\times$U(1)$_{J_2}$, which are captured by the connections $(A_{J_1},A_{J_2})$ on $Z_4$ after making the replacements 
\beq
d\hat{\phi} \to d\hat{\phi} + A_{J_1},~~~~~d \psi \to d\psi+ A_{J_2},
\eeq
in the above expressions for $\hat{A}_{1,2}$ in a gauge for ${\cal A}$ in which the Killing spinors of our $d=7$ geometries are singlets with respect to U(1)$_{J_1}\times$U(1)$_{J_2}$. The $d=2$ anomaly polynomial, in the large $N$ limit, can then be computed by fixing
\beq
c_1({\cal N}_i)= \Delta_i c_1(R_{2d})+\hat{F}_i,~~~~ \hat F_i= d\hat A_i, \label{eq: shift}
\eeq
in \eqref{eq:6danomolypolynomial} and integrating over M$_4$. Here $\Delta_1+\Delta_2=2$ to ensure that the preserved spinor has R-charge 1 and $R_{2d}$ is the pull-back of the U(1) R-symmetry bundle to $Z_4$. The central charge can then be extracted from this via c-extremisation.

Let us now perform this computation with the background gauge fields and M$_4$ identified with the wrapped M5 brane backgrounds we constructed in the previous 2 sections. 

\subsection{$\text{M}_4=\mathbb{WCP}^2_{[k,k,\ell]}$} \label{eq:QFTWCP2}
In this section we compute the central charge of the ${\cal N}=(2,0)$ SCFT in 2 dimensions that follows from compactifying the M5 brane world volume CFT on $\mathbb{WCP}^2_{[k,k,\ell]}$, the input to this computation is provided by the internal metric and gauge field ${\cal A}$ in section \ref{eq:M5braneswrappingWCP2}.

We begin by introducing coordinates $(\hat{\theta},\hat{\phi})$ that span the 2-sphere factor as in \eqref{eq:2sphereembedding}. A gauge for ${\cal A}$ in section \ref{eq:M5braneswrappingWCP2}, with respect to which spinors supported by the geometry are not charged under $(\partial_{\psi},\partial_{\hat{\phi}})$ is given by
\begin{align}
g {\cal A}&= \frac{2(1-c)(3c+1)(1-x)}{\ell c H}\left(d\psi+\frac{\ell}{2}(1-\cos \hat{\theta})d\hat{\phi}\right)+\frac{2}{\ell} d\psi+d\hat{\phi},\nn\\[2mm]
H&=1+3x+3c(1-x),~~~~~c=\frac{4k (\ell+k)}{3\ell^2}.
\end{align}
From this we quickly ascertain that the background gauge fields in the field theory, and their field strengths are given by
\begin{align}
\hat{A}_1 = \hat{A}_2 & = \alpha (d\psi + A_{J_2}) - (\eta - \beta \cos \hat{\theta}) (d \hat{\phi} + A_{J_1}),\nn \\[2mm]
\hat{F}_1 = \hat{F}_2 & = \alpha' dx \wedge (d\psi + A_{J_2} ) + \alpha F_{J_2} + (\eta' dx +\beta \sin \hat{\theta} d \hat{\theta} - \beta' \cos \hat{\theta} dx )  \wedge(d\hat{\phi} + A_{J_1} ) \nn \\[2mm]
& + (\eta - \beta \cos \hat{\theta}) F_{J_1} ,
\end{align}
where $\alpha, \beta, \eta$ are functions of $x$ defined as
\begin{equation}
\alpha = \frac{2}{\ell} \eta, \qquad \eta = \beta + \frac{1}{2}, \qquad \beta = \frac{(1-c)(3c+1)(1-x)}{2 c H},
\end{equation}
for $F_{J_i} = d A_{J_i}$ and  $c_1(J_i) = [F_{J_i}/ 2 \pi ] \in H^2 (Z_4, \mathbb{Z}^2) $. 

The anomaly polynomial in 2d is given by fixing \eqref{eq: shift} in \eqref{eq:6danomolypolynomial} for the above values of $(\hat{A}_1,\hat{A}_2)$ and their field strengths  then integrating  over the compact space $\mathbb{WCP}^2_{[k,k,\ell]}$, namely
\begin{equation}
{\cal{A}}_{2d} = \frac{\tilde{k}^2 N^3}{24} \int_{\mathbb{WCP}^2_{[k,k,\ell]}} c_1({\cal{N}}_1)^2 c_1({\cal{N}}_2)^2.
\end{equation}
Performing the integral  leads to
\begin{align}\label{eq:A6doverWCP2}
{\cal{A}}_{2d} & = \frac{\tilde{k}^2 N^3}{24} \bigg( (\Delta_1^2 + 4 \Delta_1 \Delta_2 + \Delta_2^2) I_1 c_1(R_{2d})^2 + (\Delta_1 + \Delta_2) c_1(R_{2d}) (I_2 c_1(J_1)+ I_3 c_1(J_2)) \nn \\[2mm]
& + c_1(J_1)^2 I_4 + c_1(J_1) c_1(J_2) I_5 + c_1(J_2)^2 I_6 \bigg),
\end{align}
where
\begin{align}
I_1 & = \left[ \frac{4}{\ell} \beta^2 \right]_{x=x_0}^{x=1},~~~
I_2  = \left[ \frac{4}{ \ell} \beta^2 (3+ 4 \beta)\right]_{x=x_0}^{x=1},~~~
I_3  = \frac{2}{\ell} I_2 , \\[2mm]
I_4 & = \left[ \frac{2 \beta^2}{\ell} (3 + 8 \beta + 8 \beta^2) \right]_{x=x_0}^{x=1},~~~
I_5  = \left[ \frac{8 \beta^2}{ \ell^2} (3 + 8 \beta + 6 \beta^2) \right]_{x=x_0}^{x=1}, ~~~
I_6  = \frac{1}{\ell} I_5\nn.
\end{align}
With the $d=2$ anomaly polynomial in hand it is now relatively simple to extract the holographic central charge via c-extremisation. Specifically the coefficient of $c_1(L_a)c_1(L_b)$ in $2{\cal A}_{2d}$ should be identified with $\text{tr}(\hat\gamma^{(2)} Q_a Q_b)$, for $Q_a$ the global symmetry associated to the U(1) line bundle $L_a$ over $Z_4$. c-extremisation is the statement that the U(1) R-symmetry of the $d=2$ SCFT extremises
\begin{equation}\label{eq:trialcc}
c_{\text{trial}} = 3 \text{tr} \gamma^3 R_{\text{trial}}^2.
\end{equation}
As such we write a trial R-symmetry as
\begin{equation}\label{eq:Rtrialepsilon}
R_{\text{trial}} = R_{2d} + \epsilon_1 J_1 + \epsilon_2 J_2,
\end{equation}
which allows for a mixing with the U(1)$_{J_1}\times$U(1)$_{J_2}$. Substituting this into \eqref{eq:A6doverWCP2} leads to a trial central charge of the form
\begin{equation}\label{eq:cctoextremise}
c_{\text{trial}} = - 6 \frac{\tilde{k}^2 N^3}{24} \left( (\Delta_1^2 + 4 \Delta_1 \Delta_2 + \Delta_2^2) I_1 + (\Delta_1 + \Delta_2) (I_2 \epsilon_1 + I_3 \epsilon_2) + \epsilon_1^2 I_4 + \epsilon_1 \epsilon_2 I_5 + \epsilon_2^2 I_6 \right).
\end{equation}
This must now be extremised with respect to $\epsilon_1$, $\epsilon_2$ and $\Delta_1=2-\Delta_2$, we find that this requires fixing
\begin{equation}
\epsilon_1 = 0, \qquad \epsilon_2 = - \frac{ 8 k \ell ( \ell + k)}{( \ell + 2 k)^2 + 2 \ell^2}, \qquad \Delta_1=\Delta_2 = 1.
\end{equation}
Notice this means that there is no mixing with U(1)$_{J_1}$, but as this is realising the Cartan of the SU(2) isometry on S$^2$ this was to be expected - \textit{i.e.} Abelian and non-Abelian symmetries are well known not to mix.  The central charge for the $d=2$ SCFT in the large $N$ limit is then given by fixing $(\epsilon_1,\epsilon_2,\Delta_1)$ to their extremal values in $c_{\text{trial}}$, this yields
\begin{equation}
c_{\text{CFT}} = \frac{ ( \ell - 2k )^4}{8 k^2 \ell ( (\ell+2k)^2 + 2 \ell^2 ) }  N^3 \tilde{k}^2,
\end{equation}
which exactly matches the holographic prediction computed in \eqref{eq:ccAdS3WCP2}. We also notice that by identifying $(J_1=\partial_{\hat{\phi}}, J_2=\partial_{\psi})$
\beq
\epsilon_1 J_1 + \epsilon_2 J_2  =- \frac{8 \ell k(\ell+k)}{2\ell^2+(\ell+2k)^2}\partial_{\psi},
\eeq
which reproduces the R-symmetry Killing vector of the geometry in \eqref{eq:RsymmWCP2sugra}. We thus also match the twisting of the R-symmetry. 

\subsection{$\mathbb{WCP}^1_{[k_1,k_2]}\hookrightarrow \text{M}_4\to \Sigma_{\mathfrak{g}}$ }
In this section we compute the central charges of the SCFTs dual to the 3 geometries in section \ref{sec:M5spindlesSigma}. We will be somewhat detailed for the 2-sphere, but be more brief for the other choices of the genus $\mathfrak{g}$.

We begin our analysis with the S$^2$ which we parametrise as in the previous section. For this solution, the gauge in which the spinor is a singlet with respect to $(\partial_{\psi},\partial_{\hat{\phi}})$ is 
\begin{equation}
g {\cal{A}} = \frac{2 \left(1-c^2\right) n (b+(x-2) x) }{H} \left( d \psi + \frac{(1-\cos \hat{\theta} )}{c n} d \hat{\phi} \right) + c n d \psi + d \hat{\phi}.
\end{equation}
From this we extract the following background gauge fields for the field theory 
\begin{align}
\hat{A}_1 = \hat{A}_2 = ( \eta + \beta \cos \hat{\theta} ) (d \hat{\phi} + A_{J_1} ) + \alpha ( d \psi + A_{J_2} ), 
\end{align}
where $\alpha, \beta, \eta$ are the following functions of $x$ 
\begin{equation}
\alpha = c n \eta, \qquad \eta = \frac{1}{2} - \beta, \qquad \beta = \frac{(c^2 - 1) n (b + x (x - 2))}{H c n}.
\end{equation}
By operating similar steps to what we illustrated for $\mathbb{WCP}^2_{[k,k,\ell]}$ we arrive at an expression of the same form as \eqref{eq:A6doverWCP2}, only now for integrals $I_{a}$ given by
\begin{align}
I_1 & = \left[ 2 c n \beta \right]_{x = x_1}^{x = x_2},~~~
I_2 = \left[ 2 c n \beta^2 (3 - 4 \beta) \right]_{x = x_1}^{x = x_2},~~~
I_3 = c n I_2, \\[2mm]
I_4 & = \left[ c n \beta^2 (3 - 8 \beta + 8 \beta^2 ) \right]_{x = x_1}^{x = x_2},~~~
I_5 = \left[ 2 c^2 n^2 \beta^2 (3 - 8 \beta + 6 \beta^2) \right]_{x = x_1}^{x = x_2},~~~
I_6 = \frac{c n}{2} I_5\nn,
\end{align}
with the $d=2$ anomaly in hand we can now repeat the steps performed in the previous section to arrive at the field theory central charge through c-extremisation. Taking the trial R-symmetry again as in \eqref{eq:Rtrialepsilon} to allow for potential mixing with the U(1)$_{J_1}\times$U(1)$_{J_2}$ symmetry of the compactification manifold, we find that the trial holographic charge is extremised by 
\begin{equation}
\epsilon_1 = 0, \qquad \epsilon_2 = \frac{8 k_1 k_2 ( (k_1^2 + k_1 k_2 + k_2^2)q - 3 (k_1 + k_2))}{8(k_1^2 + k_1 k_2 + k_2^2)- 3 q (k_1 + k_2)(k_1^2 + k_2^2)}, \qquad \Delta_1=\Delta_2 = 1.
\end{equation}
We again find no mixing with $J_1$, as was to be expected, and plugging these values into the trial central charge leads, at leading order in $N$, to
\begin{equation}
c_{\text{CFT}} = \frac{(k_2-k_1)^3(24 -12 (k_1+k_2) q + (k_1^2+4k_1 k_2+k_2^2) q^2 )}{8 k_1 k_2(3q(k_1+k_2)(k_1^2+k_2^2) - 8 (k_1^2+k_1 k_2+k_2^2))} N^3\tilde{k}^2 .
\end{equation}
This exactly matches the holographic central charge of \eqref{eq:hccspindlesS2}, likewise  $\epsilon_1 J_1 + \epsilon_2 J_2$ reproduce the R-symmetry Killing vector in \eqref{eq:RsymmsugraSigmakappa} with $(\kappa=1,\mathfrak{g}=0)$.\\
~\\
We now move onto the genus zero case, we will be more brief to avoid this section becoming highly redundant. We elect to take the supergravity gauge to be
\begin{equation}
g {\cal{A}} = \frac{2 n (c+(x-2) x) }{H} \left(d \psi - \frac{b }{4 n} \rho \right),
\end{equation}
for $d \rho = - \text{vol}(\mathbb{T}^2)$, which informs us what the background gauge fields should be set equal to. By implementing similar steps to before, we can define a trial central charge. This gets extremised by 
\begin{equation}
\epsilon_1 =0, \qquad \epsilon_2 = - \frac{8 k_1 k_2 (k_1^2 + k_1 k_2 + k_2^2)}{3 (k_1^3 + k_1^2 k_2 + k_1 k_2^2 + k_2^3)}.\qquad \Delta_1=\Delta_2=1.
\end{equation}
We note that once again $\epsilon_1$ is trivial, which indicates no mixing with isometries of $\mathbb{T}^2$, but there is no longer a reason to expect this a priori. Note, however, that $\epsilon_1 J_1 + \epsilon_2 J_2$ reproduces the same R-symmetry as \eqref{eq:RsymmsugraSigmakappa} for $(\kappa = 0,\mathfrak{g}=1)$, which likewise lacks such a mixing. We find the leading order central charge 
\begin{equation}
c_{\text{CFT}} = \frac{(k_2-k_1)^3(k_1^2+4k_1k_2+k_2^2)}{24 k_1 k_2(k_1+k_2)(k_1^2+k_2^2)} q \tilde{k}^2 N^3.
\end{equation}
which matches the holographic result in equation \eqref{eq:hccSpindleT2}.\\
~\\
Finally we consider the case of $\Sigma_{\mathfrak{g}}$ with $\mathfrak{g}>1$. We elect the gauge 
\begin{equation}
g {\cal{A}}= \frac{(c-1) n }{c} d \psi - \frac{2 n \left(b (c-2) c^3-x (2 c (x-2)-x+2)\right)}{c H} \left( d \psi + \frac{c \rho}{(c-1) n} \right),
\end{equation}
with $d \rho = - \text{vol}(\Sigma_{\mathfrak{g}})$. From this we extract $(\hat{A}_1,\hat{A}_2)$, compute ${\cal A}_{2d}$, then $c_{\text{trial}}$. In this case we find that c-extremisation demands
\begin{equation}
\epsilon_1 = 0, \qquad \epsilon_2 = - \frac{8 k_1 k_2 ( (k_1^2 + k_1 k_2 + k_2^2)q + 3 (k_1 + k_2)) }{8(k_1^2 + k_1 k_2 + k_2^2) + 3 q (k_1 + k_2)(k_1^2 + k_2^2)} = - \frac{2}{n}.
\end{equation}
As for the previous solutions, only $\epsilon_2$ is non trivial and $ \epsilon_1 J_1 + \epsilon_2 J_2$ reproduces \eqref{eq:RsymmsugraSigmakappa} for $\kappa = - 1$. Again there was no reason to expect no mixing with the U(1) of the Riemann surface, other than that this is consistent with the R-symmetry Killing vector of our supergravity solution. The central charge for this case is
\begin{equation}
c_{\text{CFT}} = \frac{(k_2-k_1)^3 ( 24 + 12 (k_1+k_2)q + (k_1^2+4 k_1 k_2+k_2^2)q^2)}{8 k_1 k_2 ( 3 (k_1+k_2)(k_1^2+k_2^2)q) + 8 (k_1^2+k_1 k_2+k_2^2)} (\mathfrak{g}-1) \tilde{k}^2 N^3,
\end{equation}
which reproduces the holographic prediction computed in \eqref{eq:hccSpindlesSigmag}.

In summary we find a precise matching for both central charges and R-symmetry between the field theory computations of this section and the geometric computations leading to \eqref{eq:hccSigmakappa} and \eqref{eq:RsymmsugraSigmakappa}.

\section{Conclusions}\label{sec:conclusions}

This work is concerned with constructing the first example of M5 branes wrapping the weighted projective space $\mathbb{WCP}^2$, a topological $\mathbb{CP}^2$ with 3 orbifold fixed points and 3 associated integer weights, as well as other 4 dimensional orbifolds and the two dimensional CFTs to which they are all dual. 
With this goal in mind, supersymmetric AdS$_3$ solutions of the U(1) invariant sector of minimal supergravity in 7 dimensions, which admits uplifts to $d=11$ supergravity and massive IIA, are classified, resulting in 2 classes. For the first class, which admits a point dependent scalar, the internal space is a negative curvature Kahler-Einstein manifold. For the second it is a circle fibration over a Riemann surface, all foliated over an interval. This second class is far broader, with solutions defined in terms of 3 PDEs. Focusing on the second class, an ansatz is made such that the Riemann surface has constant curvature, which reduces the 3 defining PDEs to a single second order ODE. Polynomial solutions to this ODE are found which contain both new and previously known solutions. A particular polynomial solution admits many distinct global completions, those for which the internal space is compact and free from curvature singularities are the focus of the rest of this work. The uplifts of these solutions to $d=11$ supergravity give rise to AdS$_3$ solutions that describe M5 branes wrapped on interesting 4 dimensional orbifolds. These are dual to 2 dimensional SCFTs that result from compactifying the ${\cal N}=(2,0)$ SCFT on the M5 brane world volume on these orbifolds. Studying these geometries, their uplifts and CFT duals is the  focus of the rest of the work.

Two distinct types of supersymmetric AdS$_3$ solution are constructed, these are: 1) M5 branes wrapping the weighted projective space $\mathbb{WCP}^2$. 2) M5 branes wrapping a spindle fibred over compact Riemann surfaces of arbitrary genus $\mathfrak{g}$. We compute the holographic central charges and R-symmetry Killing vectors of these solutions which provides a non-trivial prediction about their dual CFTs. Then through a field theory computation, involving the M5 brane anomaly polynomial and c extremisation, we are able to precisely match the prediction from gravity in the large $N$ limit.

There are various ways that one could build upon and/or generalise this work. An incomplete list is the following:

\begin{itemize}
\item The $d=7$ solutions we have constructed admit uplifts to both massive IIA and $d=11$ supergravity, yet we only performed a detailed study of their embeddings into $d=11$. It would be very interesting to study their embeddings into the infinite family of internal spaces for AdS$_7$ solutions in massive IIA.
\item Within the AdS$_3\times$M$_4$ solutions we have found that are defined in terms of polynomials, we only studied the global completions for which M$_4$ is compact and free from curvature singularities - this  far from exhausts the possibilities for physical solutions. There may be some low hanging fruit to be harvested by considering other global completions, see the discussion in section \ref{sec:polysols}.
\item We have classified AdS$_3$ solutions of minimal $d=7$ gauged supergravity, but one could perform a similar classification in the U(1)$^2$ invariant sector of the maximal theory - indeed this would be more natural in the context of wrapped M5 branes. This would be an interesting avenue to explore in the future.
\item Related to the last point, upon uplifting the geometries found in this work to $d=11$ it is necessary to either tune the weights of the $\mathbb{WCP}^1$ or $\mathbb{WCP}^2$ factors, or uplift in terms of S$^4/\mathbb{Z}_{\tilde{k}}$ with $\tilde{k}$ tuned. It is likely that more general forms of the solutions found in this work exist within maximal $d=7$ gauged supergravity that do not require such tunings. 
\item With M5 branes wrapped on $\mathbb{WCP}^2$ in hand, it would be worth exploring what other wrapped brane scenarios are compatible with this orbifold. In \cite{Faedo:2024upq} both M5 branes and D4 branes wrapping quadrilaterals are considered which result in AdS$_3$ and AdS$_2$ solutions respectively. This suggests that it should also be possible to construct D4 branes wrapping $\mathbb{WCP}^2$. 
\item We specifically find M5 branes wrapping the weighted projective space $\mathbb{WCP}^2_{[k,k,\ell]}$, which contains a 2-sphere, while in general $\mathbb{WCP}^2$ can have 3 entirely independent weights and does not contain a 2-sphere. That we do not find M5 branes wrapping this more general orbifold is a weakness of the constant curvature ansatz made in this work. By relaxing this assumption, solutions with M5 branes wrapping generic $\mathbb{WCP}^2_{[k_1,k_2,k_3]}$ can probably be found within the general class of section \ref{sec:class2}.
\end{itemize}

It would be interesting to explore some of these paths in the future.

\section*{Acknowledgements}
We would like to thank Alessandro Tomasiello and Matteo Kevin Crisafio for useful discussions. We acknowledge the support of the Spanish  government grants MCIU-22-PID2021-123021NB-I00, MCIU-25-PID2024-161500NB-I00 and principality of Asturias grant SV-PA-21-AYUD/2021/52177. The work of NTM is also funded by the Ram\'on y Cajal fellowship RYC2021-033794-I, AC by the Severo Ochoa fellowship PA-23-BP22-019 and DMA by the Formaci\'on de profesorado universitario fellowship FPU24-02236. DMA thanks SISSA for its kind hospitality while this work was being completed.

\appendix

\section{Recovering solutions from the literature}\label{sec:knownsols}
In this appendix we recover solutions that are already known in the literature that lie within the local polynomial solutions of \eqref{eq:CCsol1} and \eqref{eq:CCsol3}.

\subsection{Large and small ${\cal N}=(4,0)$ solutions with $d{\cal A}=0$}\label{sec:recovingtheitalians}
First we recover the two known solutions that the local polynomial solution of \eqref{eq:CCsol1} gives rise to:

One quickly realises that metric positivity in this case  requires $\kappa>0$ and that we have $d{\cal A}=0$. We fix 
\beq
\kappa=1,~~~{\cal A}=0,~~~n=1-c_1,
\eeq
without loss of generality, through a local gauge transformation and a rescaling of $\psi$. We then get two distinct solutions depending on whether or not $c_1=0$.

For $c_1=0$ we make the following redefinitions
\beq
c_2=- 64 c,~~~~~y= 8 x^2,
\eeq
and arrive at the solution
\begin{align}
g^2 ds^2&= 8x^{2/5}(c+x^4)^{2/5}\bigg[ds^2(\text{AdS}_3)+ds^2(\text{S}^3)+ \frac{x^4}{(c+x^4)^2}dx^2\bigg],~~~~X=\frac{x^\frac{4}{5}}{(c+x^4)^{\frac{1}{5}}},\\[2mm]
{\cal A}&=0,~~~~{\cal F}_4=\frac{32\sqrt{2}}{g^3}x \text{vol}(\text{AdS}_3)\wedge dx-
\frac{4\sqrt{2} x}{g^3}dx\wedge \text{vol}(\text{S}^3),
\end{align}
where we have identified 
\beq
ds^2(\text{S}^3)= \frac{1}{4} \left(ds^2(\text{S}^2)+ D\psi^2\right),
\eeq
and $\psi$ clearly has period $4\pi$. This matches the solution of \cite{Conti:2024rwd} whose uplift to massive IIA preserves small ${\cal N}=(4,0)$ supersymmetry and to $d=11$ small ${\cal N}=(4,4)$. This was originally derived in a $d=10$ context in \cite{Lozano:2022ouq} and generalises, in terms of generic $c$, a $d=7$  solution of \cite{Dibitetto:2017tve}. 

For $c_1 \neq 0$ we make the redefinitions
\beq
c_1= \frac{2}{1+\lambda},~~~~c_2=\frac{64 c}{(\lambda-1)^3 (\lambda+1)},~~~~y=\frac{8}{(\lambda-1)^2\sin^2\theta},
\eeq
upon which we find the solution takes the form
\begin{align}
 ds^2&=\frac{8}{X^2g^2\sin^2\theta} \bigg[\frac{1}{(\lambda-1)^2}ds^2(\text{AdS}_3)+\frac{1}{(\lambda+1)^2}\cos^2\theta ds^2(\text{S}^3) +X^{10} d\theta^2\bigg],\nn\\[2mm]
X^{-5}&=(\lambda-1+\sin^{-2}\theta+c \sin^2\theta )\tan^2\theta,~~~{\cal A}=0,\\[2mm]
{\cal F}_4&=\frac{16\sqrt{2}}{g^3}d\left(-\frac{1}{(\lambda-1)^3}(\lambda \sin^{-2}\theta-(\lambda c)\cos^{-2}\theta)\text{vol}(\text{AdS}_3)-\frac{1}{(\lambda+1)^3}(c\sin^2\theta+\lambda \sin^{-2}\theta)\text{vol}(\text{S}^3)\right),\nn
\end{align}
which recovers the general solution from \cite{Conti:2024rwd}, whose uplift to massive IIA preserves large ${\cal N}=(4,0)$ supersymmetry and to $d=11$ large ${\cal N}=(4,4)$, and which exhibits a regular zero at $\theta= \frac{\pi}{2}$ when $c=-\lambda$.

This exhausts the solutions contained in \eqref{eq:CCsol1}, up to taking a $\mathbb{Z}_k$ orbifold of the 3-spheres.

\subsection{Two known solutions with $d{\cal A}\neq 0$}\label{eq:knownsols2}
We now recover the two known solutions that the local polynomial solution of \eqref{eq:CCsol3} gives rise to:\\

The first solution follows by tuning $c_1=\kappa$, where it is quick to see that a positive metric requires  $\kappa<0$, so we fix
\beq
\kappa=-1,
\eeq
without loss of generality. It also follows that
\beq
{\cal B}=0~~~\Rightarrow  D\psi=d\psi.
\eeq
Making the redefinitions
\beq
c_2=-8 a,~~~~ y=2x,
\eeq
we are lead to the solution of the local form
\begin{align}
ds^2&=\frac{1}{g^2}\left(\frac{3^8}{2}\right)^{\frac{1}{5}}\bigg[\frac{4}{9}x ds^2(\text{AdS}_3)+\frac{x dx^2}{\tilde{q}}+\frac{n^2 \tilde{q}}{36 x^2}d\psi^2+\frac{1}{3}ds^2(\mathbb{H}^2)\bigg],~~~~\tilde{q}=4x^3-9 x^2+6a x-a^2, \nn\\[2mm]
X&=\left(\frac{3}{4}\right)^{\frac{1}{5}},~~~~g{\cal A}=n\frac{a-3x}{2x}d\psi+\rho,~~~d\rho=-\text{vol}(\mathbb{H}^2),~~~~{\cal F}_4= \frac{a n}{\sqrt{2}g^3x^2}d\psi\wedge dx\wedge \text{vol}(\mathbb{H}^2),
\end{align}
we see that the metric is a direct product of the $d=5$ AdS$_3\times \mathbb{WCP}^1$ from \cite{Ferrero:2020laf}  and $\mathbb{H}^2$ - the full $d=7$ solutions was found recently in section 3.4 of \cite{Conti:2025wyj}.

The second solution follows from the tuning $c_1=0$, which again requires that $\kappa<0$. This time fixing
\beq
\kappa=-1,~~~~c_2=-\frac{64}{27}a,~~~~n=\frac{3}{2},~~~~y= \frac{8 x}{9},
\eeq
we find the solution becomes
\begin{align}
ds^2&=\frac{8}{9 g^2}x^{\frac{1}{5}}(a+x^2)^{\frac{2}{5}}\bigg[ds^2(\text{AdS}_3)+\frac{9}{16 Q}x dx^2+\frac{9}{4(a+x^2)^2}Q D\psi^2+\frac{3}{4} ds^2(\mathbb{H}^2)\bigg],\nn\\[2mm]
g {\cal A}&=-\frac{2 x^2}{a+x^2}D\psi,~~~~ D\psi=d\psi-\frac{2}{3}\rho,~~~~d\rho=-\text{vol}(\mathbb{H}^2),~~~~X= \frac{x^{\frac{2}{5}}}{(a+x^2)^{\frac{1}{5}}}, \\[2mm]
{\cal F}_4&=\frac{16\sqrt{2}}{27g^3}\text{vol}(\text{AdS}_3)\wedge dx-\frac{\sqrt{2}}{3g^3} dx\wedge d\psi\wedge \text{vol}(\mathbb{H}^2), ~~~~ Q=-x^3+\frac{1}{4}(a+x^2)^2,\nn 
\end{align}
which is locally the $q_1=q_2=a$ limit of the embedding of the AdS$_3\times\mathbb{H}^2$ vacua of minimal $d=5$ gauged supergravity into $d=7$ maximal gauged supergravity, as found in section 4 of \cite{Cheung:2022ilc}.

The above solutions in no way exhaust the global completions that \eqref{eq:CCsol3} can give rise to, neither for that matter do the explicit new solutions we present in the main text. 

\section{Relation between $d=10$ and $d=11$ uplifts}\label{sec:uplift11d}
These $7d$ solutions can be uplifted to massless type IIA and from there to $11d$. We will perform the uplift to $10d$ following the consistent truncation described in \cite{Passias:2015gya}. This procedure will uplift the solution to a class of solutions over a deformed $S^2$ over an interval, specified by a function $\alpha(z)$ with $z$ being the parametrization of the interval. The uplifted metric looks like:
\begin{equation}
\frac{ds_{10}^2}{\sqrt{2\pi}}=g^2\sqrt{-\frac{\alpha}{\alpha''}}X^{-1/2}\,ds_7^2 + X^{5/2}\sqrt{-\frac{\alpha''}{\alpha}}\left(dz^2 + \frac{\alpha^2}{\alpha'^2 - 2\alpha\alpha'' X^5} Ds^2(\text{S}^2)\right),
\end{equation}
where $Ds^2(\text{S}^2)$ is given by
\begin{equation}
Ds^2(\text{S}^2)=d\theta^2 + \sin^2\theta (d\phi - g {\cal{A}})^2,
\end{equation}
and the volume form of this deformed S$^2$ is $\text{vol}_2=\sin{\theta}d\theta\wedge(d\phi - g\cal{A})$. The function $\alpha(z)$ will satisfy the next Bianchi Identity dependent on the Romans mass:
\begin{equation}
\alpha(z)'''=-2\pi^3 3^4 F_0,
\end{equation}
but since we are interested only in the massless case ($F_0=0$) to be able to uplift the solutions to $11d$, this condition reads $\alpha'''=0$. The function $\alpha(z)$ is then a second degree polynomial $\alpha(z)=a_2 z^2+a_1 z +a_0$, which will allow to simplify a lot the expressions for the metric by using difeomorphism invariance. First we can eliminate the linear term in $z$ through the shift $z\xrightarrow{}z-a_1/2a_2$ and then choose $a_2$ to our convenience through rescaling $z\xrightarrow{}\lambda z$, ending with
\begin{equation}
\alpha(z)=\frac{81 \tilde{k} \pi^2}{2 \xi}(z^2-1),
\end{equation}
with $\tilde{k}$ being the only undetermined constant. To further simplify the metric we will change from the variable $z$ to the variable $\beta$ as $z=\sin{\beta}$. The dilaton and the fluxes in $10d$ are
\begin{align}
e^\Phi&=\frac{\sqrt{2\pi \xi}}{\tilde{k}}\frac{\cos^{3/2}\beta}{\sqrt{\Delta X^{3/2}}},\nn\\[2mm]
F_2&=d C_1 = - \frac{\tilde{k}}{2}\left(\text{vol}_{2} + g\cos{\theta}\mathcal{F}\right),\nn\\[2mm]
F_4&=\frac{g \xi \tilde{k} \pi}{2}\bigg( \sin^2{\theta}\cos\beta d\beta\wedge\mathcal{F}\wedge (d\phi-g\mathcal{A}) -\frac{\cos^2\beta\sin\beta}{2X^4\Delta}\cos{\theta}\mathcal{F} \wedge \text{vol}_{2} \nn\\[2mm]
& + 2g X^4\cos\beta d\beta\wedge\star_7\mathcal{F}_4-\sqrt{2}g^2\sin\beta\mathcal{F}_4\bigg), \\[2mm]
H&=dB_2,~~~~B_2=\pi \xi \left(\frac{\cos^2\beta\sin\beta}{2\Delta X^4}\text{vol}_{2} +\cos{\theta}\cos\beta d\beta \wedge(d\phi - g \mathcal{A})\right), \nn
\end{align}
where $\Delta=X^{-4}\sin^2\beta+X\cos^2\beta$. The uplift of the $10d$ metric to $11d$ looks
\begin{equation}
ds_{11}^2=e^{-\frac{2}{3}\phi}ds_{10}^2+e^{\frac{4}{3}\phi}(d \hat{\psi} - C_1)^2,~~~~C_1=\frac{\tilde{k}}{2}\cos\theta (d\phi-g {\cal A}).
\end{equation}
Substituting the values for the dilaton we arrive at
\begin{align}
\frac{ds_{11}^2}{L^2} = & g^2\Delta^{1/3}g_{\mu\nu}dx^{\mu}dx^{\nu}+2\Delta^{-2/3}ds^2(\text{B}_4),\nn\\[2mm]
ds^2(\text{B}_4) = & X^3\Delta d\beta^2+\frac{\cos^2{\beta}}{X}\left(\frac{1}{4}Ds^2(\text{S}^2)+\frac{1}{\tilde{k}^2}(d\hat{\psi} - C_1)^2\right), \label{eq:11dmets}
\end{align}
where $L=\left(\frac{\tilde{k} \xi \pi}{\sqrt{2}}\right)^{\frac{1}{3}}$ is just a collection of constants. The $11d$ flux takes the form
\begin{align}
G &= F_4 + (d\hat{\psi} - C_1)\wedge H_3,\nn\\[2mm]
\frac{G}{L^3} & = \frac{g}{\sqrt{2}} \bigg( \sin^2\theta\cos\beta d\beta\wedge\mathcal{F}\wedge(d\phi-g\mathcal{A}) -\frac{\cos^2\beta\sin\beta}{2 X^4 \Delta}\cos\theta\mathcal{F}\wedge\text{vol}_2 -\sqrt{2} g^2 \sin\beta\mathcal{F}_4 \nn\\[2mm]
& + 2 g X^4 \cos\beta d\beta\wedge\star_7\mathcal{F}_4 \bigg) + \frac{\sqrt{2}}{\tilde{k}} (d \hat{\psi} - C_1) \wedge\bigg(g\cos\beta\cos\theta d\beta\wedge\mathcal{F} \\[2mm]
& + \frac{g\cos^2\beta\sin\beta}{2 X^4 \Delta} \sin\theta d\theta \wedge \mathcal{F} - \frac{5\cos^4\beta\sin\beta }{2 X^4 \Delta^2} dX \wedge \text{vol}_2 + \frac{(2X^5-1) \Delta + 2 X }{2 X^4 \Delta^2} \cos^3\beta d\beta\wedge\text{vol}_{2} \bigg).\nn
\end{align}
with $H_3=dB_2$. The flux and the metric fulfil the equations of motion in 11d
\begin{align}
R_{\mu\nu} - \frac{1}{12}(G_{\mu\nu}-\frac{1}{12} G^2 g_{\mu\nu})&=0, \\[2mm]
d \star G + \frac{1}{2}G \wedge G &=0. \nn 
\end{align}

\section{M5 branes wrapping a simple $\tilde{\Sigma}\times\Sigma_{\mathfrak{g}}$ with $\mathfrak{g}>1$ but fixed}\label{sec:simple H_2}
The simplest of our solutions with spindles fibred over $\Sigma_{\mathfrak{g}}$ shares the same tuning of $c_2$ as the $\mathbb{WCP}^2$ solution in section \ref{eq:M5braneswrappingWCP2}, namely
\beq
c_2= \frac{4 c_1^3(c_1-2\kappa)}{(c_1-\kappa)^3}.\label{eq:c2tunign}
\eeq
For this it is also possible to define a global solution containing a spindle. If we redefine
\beq
y=\frac{2x c_1^2}{(c_1-\kappa)^2},
\eeq
we once more have that the internal space vanishes as $\mathbb{R}^4$ at $x=1$, but this time  we  would like to tune $c_1$ such that $x=1$ lies outside the domain of $x$. It does not take long to realise that a well defined metric only allows this for $\kappa<0$ and $2\kappa<c_1$. It is convenient to tune
\beq
\kappa=-1,~~~~~c_1=-2- \frac{3}{8}\frac{(k_2-k_1)}{8 k_1 k_2},~~~~n=\frac{3k_1^2+2k_1 k_2+3 k_2^2}{4 k_1 k_2(k_1+k_2)},~~~~k_2>k_1>0,
\eeq
for $k_i$ integer, and then make a local gauge transformation such that the gauge field and phase in the canonical vielbein direction $W$ are mapped to
\beq
g{\cal A}\to g {\cal A}+ \frac{(k_1-k_2)(k_2-k_1+4(k_2^2 m_1+ k_2^2 m_1))}{4 k_1 k_2 (k_1+k_2)}d\psi,~~~~~ e^{i n\psi}\to e^{i(m_2-m_1)\psi},
\eeq
where $m_1,m_2$ are integers such that $k_1 m_2-k_2 m_1=1$, which always exist if $(k_1,k_2)$ share no common prime factor due to  B\'ezout's identity - we are thus further assuming that
\beq
\text{gcd}(k_1,k_2)=1,~~~~ k_1 m_2-k_2 m_1=1,
\eeq
this ensures that the spinors the background support transform properly as the $\partial_{\psi}$ circle is traversed, provided $\psi\in[0,2\pi)$. 

The resulting solution then takes the form
\begin{align}\label{eq:AdS3SpindleSigmag}
ds^2&=\frac{2(3k_1+k_2)^2(k_1+3k_2)^2x}{g^2X^2} \bigg[\frac{1}{\lambda}ds^2(\text{AdS}_3)+ds^2(\tilde{\Sigma})+ \frac{1-x}{64 k_1 k_2(k_1+k_2)^2 x }ds^2(\Sigma_{\mathfrak{g}})\bigg] ,\nn\\[2mm]
ds^2(\tilde{\Sigma})&=\frac{dx^2}{(1-x)Q_1Q_2}+ \frac{(1-x)Q_1 Q_2}{64 k_1^2k_2^2(k_1+k_2)^2 x Q_3^2}D\psi^2,~~~~D\psi=d\psi-\frac{k_1+k_2}{2}\rho\nn\\[2mm]
g{\cal A}&= \frac{(k_1-k_2)m_1}{k_1} d \psi -\frac{(3k_1+k_2) Q_1}{2k_1k_2 Q_3}d\psi-\frac{(k_2-k_1)^2(3k_1+k_2)(k_1+3k_2)(1-x)}{4 k_1 k_2 Q_3}\rho,\nn\\[2mm]
{\cal F}_4&=-\frac{(k_1-k_2)^2(3k_1+k_2)^2(k_1+3k_2)^2}{g^3}\bigg[\frac{2\sqrt{2}}{\lambda^3}\text{vol}(\text{AdS}_3)\wedge dx\nn\\[2mm]
&+ \frac{(1-x^2)}{128\sqrt{2}k_1^2k_2^2(k_1+k_2)^3 x^2}dx\wedge d\psi\wedge \text{vol}(\Sigma_{\mathfrak{g}}) \bigg],~~~~X^5=\frac{(3k_1+k_2)(k_1+3k_2)x}{Q_3},
\end{align} 
where we define
\begin{align}
Q_1&=(k_1+3k_2)^2x-(k_1-k_2)^2,~~~~Q_2=(k_1-k_2)^2-(3k_1+k_2)^2x\nn\\[2mm]
Q_3&=(k_1-k_2)^2+(3k_1+k_2)(k_1+3k_2)x,~~~~\lambda=3 k_1^2+2 k_1k_2+3 k_2^2
\end{align}
and we have taken a quotient such that  $\mathbb{H}^2\to \mathbb{H}^2/\Gamma=\Sigma_{\mathfrak{g}}$ a genus $\mathfrak{g}$ Riemann surface with $\mathfrak{g}>1$  as in \eqref{eq:H2quotient}. The $x$ coordinate is bounded as $x_1\leq x\leq x_2<1$ for 
\beq
x_1= \frac{(k_1-k_2)^2}{(k_1+3 k_2)^2},~~~~x_2= \frac{(k_1-k_2)^2}{(3k_1+ k_2)^2},
\eeq
with $Q_i$ for $i=1,2$ positive between these bounds and vanishing at $x=x_i$ and $Q_3$ positive for the whole interval.
We find that as  $x\to x_i$
\beq
ds^2(\tilde{\Sigma})\propto dz^2+ \frac{z^2}{k_i^2}D\psi^2,~~~~{\cal A}\to  \frac{k_1-k_2}{k_i}m_i d\psi- \left( 1 + \frac{k_1 + k_2 }{2 k_i} \right)\rho,
\eeq
where $x= x_i \pm  z^2$ as appropriate and the rest of the metric is constant. We thus see that  $\tilde{\Sigma}$ has the topology of a spindle when $\psi\sim \psi+2\pi$ and ${\cal A}$ is behaving as a gauge field with support on such a space should at the $\mathbb{R}^2/\mathbb{Z}_{k_i}$ fixed points, it is not hard to show that
\beq
\frac{g}{2\pi}\int_{\tilde{\Sigma}}d{\cal A} =  \frac{k_1 - k_2}{k_1k_2},\label{eq:fluxcharge1}
\eeq
as befitting the anti-twist preservation of supersymmetry on the spindle. We also confirm that the Euler characteristic on $\tilde{\Sigma}$ is as expected, \textit{i.e.} upon sending $D\psi\to d\psi$ we find
\begin{equation}
\chi_E(\tilde{\Sigma}) = \frac{1}{4 \pi} \int_{\tilde{\Sigma}} R \text{vol}(\tilde{\Sigma}) = \frac{k_1 + k_2}{k_1 k_2} = 2 - \left( 1- \frac{1}{k_1} \right) - \left( 1 - \frac{1}{k_2}\right),
\end{equation}
As $\tilde{\Sigma}$ is fibred over $\Sigma_{\mathfrak{g}}$ in the full space, we should have that
\beq
\frac{1}{2\pi}\int_{\Sigma_{\mathfrak{g}}}dD\psi = (\mathfrak{g}-1)(k_1+k_2),
\eeq 
is an integer, which in this case is automatic. On the other hand, upon lifting to $d=11$ we have that ${\cal A}$ is a connection term which fibres S$^4$ over the metric in \eqref{eq:AdS3SpindleSigmag} --  as such we should also have that
\beq
\frac{g}{2\pi}\int_{\Sigma_{\mathfrak{g},x=x_i}}d{\cal A}\in\mathbb{Z},
\eeq
so that this fibration is well defined. This demands that we tune the genus as
\beq
\mathfrak{g}=1 + k_1 k_2 q,~~~q\in\mathbb{N}
\eeq
We thus find that the various connections give rise to the charges \eqref{eq:fluxcharge1} and
\beq
\frac{1}{2\pi}\int_{\Sigma_{\mathfrak{g}}}dD\psi= q k_1 k_2(k_1+k_2) ,~~~\frac{g}{2\pi}\int_{\Sigma_{\mathfrak{g},x=x_1}}d{\cal A}=qk_2(3k_1+k_2),~~~~ \frac{g}{2\pi}\int_{\Sigma_{\mathfrak{g},x=x_2}}d{\cal A}= qk_1(k_1+3k_2),
\eeq
when the geometry is well defined. As in the other case, lifting to $d=11$ in terms of S$^4/\mathbb{Z}_{\tilde{k}}$ requires that either $\tilde{k}$ is even, or $q$ is even and $(k_1,k_2)$ are odd. Flux quantisation for the $d=11$ 4-form over S$^4/\mathbb{Z}_{\tilde{k}}$ also fixes $L$ as in \eqref{eq:Lfixed}, and we find that at $\sin\beta=\pm 1$ we can define the charges
\beq
\mp \frac{1}{(2\pi)^3}\int_{\tilde{\Sigma}\times \Sigma_{\mathfrak{g}}}G= \frac{(k_2-k_1)(k_1^2+6 k_1 k_2 +k_2^2)}{8k_1 k_2}\tilde{k}N q,
\eeq
which is again rational but may be made integer by tuning for instance $N$. We can then compute the central charge of this solution through the holographic formula \eqref{eq:holcent}, we find
\beq
c_{\text{hol}}= \frac{q(k_2-k_1)^3(k_1^2+k_2^2+14k_1 k_2)}{8 k_1 k_2(3(k_1^2+k_2^2)+2 k_1 k_2)}N^3 \tilde{k}^2.
\eeq
While this solution has the same topology as that of section 4 of \cite{Cheung:2022ilc}, a tuning of which is recovered in section \ref{eq:knownsols2} (after replacing $\mathbb{H}^2$ with $\Sigma_{\mathfrak{g}}$), one needs only to compare the warping of the AdS$_3$ and $\Sigma_{\mathfrak{g}}$ factors of the metric to see that they are distinct. Indeed due to this warping and unlike the solution of \cite{Cheung:2022ilc}, this solution cannot arise from the embedding of some minimal $d=5$ supergravity solution into $d=7$.

\printbibliography

\end{document}